\documentclass[journal]{IEEEtran}

\usepackage{cite,graphicx,amsmath,amssymb}
\usepackage{color,psfrag}

\usepackage{citesort}
\usepackage{comment}
\usepackage{cases}
\usepackage{accents}
\usepackage{stfloats}
\usepackage{times}
\usepackage{relsize}
\usepackage{tabularx}
\usepackage{relsize}
\usepackage{epstopdf}
\usepackage{multirow}
\usepackage{subcaption}
\usepackage[font=scriptsize]{caption}
\makeatletter
\renewcommand{\maketag@@@}[1]{\hbox{\m@th\normalsize\normalfont#1}}%
\makeatother

\renewcommand{\QED}{\QEDopen}
\newtheorem{theorem}{Theorem}
\newtheorem{lemma}{Lemma}
\newtheorem{corollary}{Corollary}

\makeatletter
\newcommand{\figcaption}{\def\@captype{figure}\caption}
\newcommand{\tabcaption}{\def\@captype{table}\caption}
\makeatother

\newcommand{\bs}{\boldsymbol}

\newcommand{\bOm}{\boldsymbol{\Omega}}
\newcommand{\bbe}{\mathbb{E}}
\newcommand{\bbc}{\mathbb{C}}

\newcommand{\bA}{\mathbf{A}}
\newcommand{\bB}{\mathbf{B}}
\newcommand{\bH}{\mathbf{H}}

\newcommand{\bG}{\mathbf{G}}
\newcommand{\bg}{\mathbf{g}}
\newcommand{\ba}{\mathbf{a}}
\newcommand{\bx}{\mathbf{x}}
\newcommand{\bb}{\mathbf{b}}

\newcommand{\nsp}{\negthickspace}
\newcommand{\nmsp}{\negmedspace}

\begin{document}
\title{Power Allocation Schemes for Multicell Massive MIMO Systems}
\author{Qi~Zhang,
        Shi~Jin,~\IEEEmembership{Member,~IEEE,}
        Matthew~McKay,~\IEEEmembership{Senior Member,~IEEE,} \\
        David~Morales-Jimenez,~\IEEEmembership{Member,~IEEE,}
        and~Hongbo Zhu% <-this % stops a space
\thanks{%Paper approved by  A. Lee Swindlehurst, the Editor for IEEE Journal of Selected Topics in Signal Processing.
        Manuscript received November 15, 2014; revised March 30, 2015.
        This work was partly supported by the China 973 project under Grant 2013CB329005, the China 863 Program under Grant 2014AA01A705 and the Jiangsu Graduate Research and Innovation Project under Grant KYLX$\_$0806. The work of S. Jin was supported by the National Natural Science Foundation of China under Grant 61222102 and the International Science ${\&}$ Technology Cooperation Program of China under Grant 2014DFT10300. The work of M. McKay and D. Morales-Jimenez was supported by the Hong Kong Research Grants Council under grant number 616713.}
%\thanks{Copyright (c) 2015 IEEE. Personal use of this material is permitted. However, permission to use this material for any other purposes must be obtained from the IEEE by sending a request to pubs-permissions@ieee.org.}
\thanks{Q. Zhang and H. Zhu are with Jiangsu Key Laboratory of Wireless Communications, Nanjing University of Posts and Telecommunications, Nanjing, 210003, P. R. China (email: zhangqiqi$\_$1212@126.com; zhuhb@njupt.edu.cn).}
\thanks{S. Jin is with the National Mobile Communications Research Laboratory,
Southeast University, Nanjing, 210096, P. R. China (email: jinshi@seu.edu.cn). S. Jin is the corresponding author.}
\thanks{M. McKay and D. Morales-Jimenez are with the Department of Electronic and Computer Engineering, Hong Kong University of Science and Technology, Clear Water Bay, Kowloon, Hong Kong (email:eemckay@ust.hk; eedmorales@ust.hk).}  }

\markboth{IEEE TRANSACTIONS ON WIRELESS COMMUNICATIONS,~Vol.~x,
No.~xx,~xx~201x} {IEEE TRANSACTIONS ON WIRELESS COMMUNICATIONS,~Vol.~x,
No.~xx,~xx~201x}

\maketitle

\begin{abstract}
This paper investigates the sum-rate gains brought by power allocation strategies in multicell massive multiple-input multiple-output systems, assuming time-division duplex transmission. For both uplink and downlink, we derive tractable expressions for the achievable rate with zero-forcing receivers and precoders respectively. %which hold for any number of base station antennas ($M$).
%we derive tractable lower and upper bound for the achievable rate with zero-forcing (ZF) receivers and a more tight approximation which lies between these two bounds and holds for any number of base station (BS) antennas ($M$).
%got directly from Jensen's inequality
%For downlink, we also give a tractable lower bound for the achievable rate with ZF precoders, which applies to arbitrary $M$ and is shown to be very tight as well.
%Based on these analytical results, the corresponding transmit power allocation schemes are given to maximize the sum rate per cell and it is found that, compared to the equal power allocation, these schemes bring considerable gains on the rate performance, especially for low or moderate (practical) numbers of antennas up to a few hundreds. When the number of BS antennas grows, these improvements abate and in the asymptotic regime of $M \to \infty$, the equal power allocation policy becomes asymptotically optimal.
{To avoid high complexity joint optimization across the network, we propose a scheduling mechanism for power allocation, where in a single time slot, only cells that do not interfere with each other adjust their transmit powers. Based on this, corresponding transmit power allocation strategies are derived, aimed at maximizing the sum rate per-cell.} These schemes are shown to bring considerable gains over equal power allocation for practical antenna configurations (e.g., up to a few hundred). However, with fixed number of users ($N$), these gains diminish as $M \to \infty$, and equal power allocation becomes optimal.
%For the more general case where $N$ can also change with $M$, we involve a parameter $M/N$ratio to do the analysis and find that with fixed ratios, the power allocation gains grow as $M$ increases.
A different conclusion is drawn for the case where both $M$ and $N$ grow large together, in which case: (i) improved rates are achieved as $M$ grows with fixed $M/N$ ratio, and (ii) the relative gains over the equal power allocation diminish as $M/N$ grows.
Moreover, we also provide applicable values of $M/N$ under an acceptable power allocation gain threshold, which can be used as
%, which can be used to determine when the power allocation schemes are worth. All these observations give us the conclusion that for practical massive MIMO systems, our power allocation schemes still have a good performance.
%Based on these results, we then provide
%practical design rules of thumb to determine appropriateness of our power allocation schemes over the equal assignment policy. These results reveal the applicability of our allocation schemes under a wide range of scenarios with practical numbers of users and antennas.
to determine when the proposed power allocation schemes yield appreciable gains, and when they do not. {From the network point of view, the proposed scheduling approach can achieve almost the same performance as the joint power allocation after one scheduling round, with much reduced complexity.}
\end{abstract}

%\newpage

\section{Introduction}\label{sec: introduction}
Multiple-input multiple-output (MIMO) systems are now regarded as a key technology for cellular communication systems due to various advantages they offer over single-antenna systems, including transmit diversity, high data rates and reliability \cite{paulraj03,tse05,jafarkhani05}. By transmitting parallel data streams, it is known that the ergodic capacity of MIMO channels increases linearly with the minimum number of transmit and receive antennas \cite{telatar99}. The transmitted information streams are separated at the receiver by means of appropriate signal processing techniques. The optimum receivers and precoders lead to a complexity burden on the system implementation. Therefore, linear receivers and precoders such as those based on the zero-forcing (ZF) criteria are often considered, as they offer significantly lower complexity with tolerable performance \cite{chen07}.

Recently, multiuser MIMO (MU-MIMO) systems, in which the antenna array of the base station (BS) simultaneously serves a multiplicity of autonomous users in the same time-frequency resource, have attracted substantial interest. Compared with conventional MIMO systems, MU-MIMO can offer a spatial multiplexing gain even if the users have only a single antenna
\cite{caire03,viswanath03,Gesbert07,caire10,yoo06}.
%ULTIPLE-INPUT multiple-output (MIMO) systems
%have attracted significant research interest during the
%last decade due to various advantages they promise, both
%in single user [1] and multiuser channels [2]. It has been
%recently shown that the employment of an excess of antennas
%at the BS (very large MIMO) offers unprecedented array and
%multiplexing gains both in the uplink and in the downlink
%[3], [4].

Though MU-MIMO has been supported in 3GPP standards, it can not achieve the transmission rates demanded in 5G. Therefore, the concept of massive MIMO, which considers the use of hundreds of antenna elements to serve tens of users simultaneously, has come to the forefront of wireless communications research \cite{rusek13}. Massive MIMO can reap all the benefits of conventional MIMO at a greater scale,
%and With large antenna arrays, the effect of fast channel fading can be averaged out based on the law of large numbers \cite{telatar99} and
%recently, lots of efforts have been made to explore the spectral and energy efficiency of massive MIMO systems from extensive perspectives
whilst providing substantial improvements in energy and spectral efficiencies with low complexity
\cite{marzetta06,marzetta10,Hoydis13,Ngo11,pitarokoilis12,wagner10,ngo13multicell,yin13,fernandes13,jose11,appaiah10,qi14}. In particular, if the number of BS antennas is far greater than the number of  single-antenna users, the simplest linear receivers and precoders become optimal \cite{Hoydis13}. It is also revealed in \cite{marzetta10} that, when the number of BS antennas grows without bound, the achievable rate can be improved to a considerable level and the uncorrelated noise, fast fading and intracell interference all vanish. Moreover, \cite{Ngo11} demonstrates that the transmitted power in massive MIMO systems can be proportionally scaled down with the number of BS antennas while maintaining the same rate performance.

%Most previous massive MIMO work assumes the equal transmit power allocation. However,
In power-limited systems, the equal power allocation policy cannot take full advantage of the energy resource and it is far from optimal. In this paper, allowing for different transmit powers at each user, we derive three tractable expressions for the achievable uplink rate---a lower bound, an upper bound and an approximation which lies between these two bounds. {To avoid the joint optimization across the network, which involves high complexity algorithms with (potential) convergence issues, we proposed a scheduling mechanism for power allocation, where in a single time slot, only cells that do not interfere with each other adjust their
transmit powers.} Based on this, we subsequently derive power allocation strategies which optimize each of these expressions and compare their relative performance.
%the lower and upper bound for the achievable uplink rate and a more tight approximation, which lies between these two bounds.
In addition, we present a tractable lower bound for the achievable downlink rate which is shown to be very tight, and similarly, a corresponding power allocation policy is derived. These schemes are shown to yield noticeable rate improvements over equal power allocation under various conditions. This is true, for example, for BSs housing up to a few hundred antennas ($M$). As $M$ increases, the gain brought by optimized power allocation becomes less significant, with the optimal assignment approaching uniform allocation as $M \to \infty$ for fixed numbers of users ($N$).
%Through these results,
%it is found that compared to equal power assignment, these allocation schemes provide noticeable rate improvements and these improvements are related to the number of BS antennas ($M$). For antenna arrays comprising up to a few hundred antennas, the proposed power allocation schemes have a demonstrable effect on the rate performance. As $M$ increases with a fixed number of users ($N$), this effect becomes less obvious and it is found that, when $M \to \infty$, the power allocation maximizing the sum rate approaches an equal assignment.
To further characterize these effects more precisely, we also consider the regime where $M$ and $N$ grow large together, i.e., with $M/N$ fixed, and show that: (i) improved rates are achieved as $M$ and $N$ both grow, and (ii) the relative gains over the equal power allocation depend on the ratio $M/N$, with larger gains for smaller ratios.
By specifying a minimum required gain over the equal power assignment, we then provide the range of $M/N$ ratios which satisfy the proposed constraint, and these results can be used to provide engineering insight into when the gains brought by optimized power allocation are sufficient to warrant the additional complexity relative to equal-power allocation, and when they are not. For example, for the uplink, if the total transmit power per cell is $20$dB, systems with $M/N<12$ (i.e., less than $12$ BS antennas per user) can obtain more than $10\%$ rate gain with optimized power allocation.
%and for the downlink, if the total transmit power per cell is $40$dB, systems with $M=100$ can also obtain more than $10\%$ rate gain with more than $10$ users.
These observations indicate that while equal power allocation is asymptotically optimal in the theoretical massive MIMO regime (i.e., when $M/N \to \infty$), for a wide range of practical scenarios in which the BS antennas is moderately larger than the number of users (but not infinitely larger), optimized power allocation can bring appreciable gains.

% Moreover, we also find that the gains brought by the proposed power allocation schemes get larger as the number of cells increase.
%In particular, these improvements also reduce  as $M$ increases, which implies that in massive MIMO systems, the power allocation maximizing the sum rate approaches an equal assignment.

This paper is organized as follows. Section \ref{sec: system model} describes the multicell MIMO system model for both uplink and downlink, and defines the associated achievable rate with ZF receivers and ZF precoders, respectively. In Section \ref{sec: uplink}, we present the lower and upper bound for the achievable uplink rate, followed by a more accurate approximation which lies between these bounds. Based on this approximation, the corresponding power allocation scheme is put forth. In Section \ref{sec: downlink}, we analyze the achievable downlink rate and, based on a tractable lower bound which we derive, an effective power allocation scheme is proposed. In Section \ref{sec: numeric results}, we provide a set of numerical results, while Section \ref{sec: conclusion} summarizes the main results of this paper.

{\em Notation}---Throughout the paper, vectors are expressed in lowercase boldface letters while matrices are denoted by uppercase boldface letters. We use ${{\bf{X}}^H}, {{\bf{X}}^T}, {\bf{X}^*}$ and $\mathbf{X}^{-1}$ to denote the conjugate-transpose, transpose, conjugate and inverse of $\bf{X}$, respectively. Moreover, $\mathbf{I}_N$ denotes an $N\times N$ identity matrix, and $[{\bf{X}}]_{ij}$ is the ($i,j$)th entry of $\bf{X}$. Finally, $\mathbb{E}\left\{  \cdot  \right\}$ is the expectation operator, $\left\| {\, \cdot \,} \right\|$ is the Euclidean norm, $(a)^+$ denotes $\max\left\{a,0\right\}$ and ${\tt tr}\left( \cdot \right)$ is the trace operation.

\section{System model}\label{sec: system model}
%\begin{figure}
%\centering{\includegraphics[scale=0.4]{multicell.eps}}
%\caption{The multicell MIMO system with $L$ cells and each cell has a $M$-antenna BS and $N$ single-antenna users.}\label{fig 0}
%%\source{For  high temperatures (dashed line) the Coulomb blockade remains}
%%\source{For lower temperatures (solid line) the Coulomb blockade is overcome}}\label{kondodotresistance}
%\setlength{\abovecaptionskip}{0pt}
%\setlength{\belowcaptionskip}{0pt}
%\end{figure}
%As shown in Fig. \ref{fig 0},
 {Consider a multicell scenario with a central cell, denoted as cell $0$, and $L$ interference cells surrounding it, denoted as cell $1$ to $L$.} Each cell contains a MU-MIMO system with $N$ single-antenna users and one $M$-antenna BS.
%We consider the uplink transmission and users transmit signals to their BSs in the same time-frequency resource.
Channel reciprocity is exploited between uplink and downlink.
\subsection{Uplink}
 On the uplink, it is assumed that all users simultaneously transmit data streams to their BSs in the same time-frequency resource and the transmit power for each user may be different. {Let $\mathbf{G}_{il}~(i,l=0,1,\ldots,L)$ denote the $M \times N$ MIMO channel matrix between the $N$ users in the $l$th cell and the $M$ BS antennas in the $i$th BS. Therefore, the $M \times 1$ received vector at the BS of the central cell can be written as %\cite{ngo13multicell}
\begin{equation}\label{system model}
\mathbf{y}_0 ^{\tt ul}= \bG_{00}\bOm_0^{1/2} \bx_0 + \sum_{l=1 }^L  {\bG_{0l}\bOm_l^{1/2} \bx_l} + \mathbf{n}_0^{\tt ul},
\end{equation}
where the $N \times N$ diagonal matrix $\bOm_l$ contains $\left[p_{1l}^{\tt ul},\ldots,p_{nl}^{\tt ul},\ldots,p_{Nl}^{\tt ul}\right]$ along its main diagonal, while $p_{nl}^{\tt ul}$ is the transmitted power of the $n$th user in the $l$th cell, $\bx_l \in \bbc^{N \times 1}$ denotes the transmitted signal vector from all users in the $l$th cell, and $\mathbf{n}_0^{\tt ul} \in \bbc^{M \times 1}$ represents the vector of additive white Gaussian noise with entries having zero mean and unit variance.}

We denote the independent channel coefficient between the $n$th user in the $l$th cell and the $m$th antenna of the $i$th BS as ${g_{minl}} = {[\mathbf{G}_{il}]_{mn}}$, which
accounts for independent fast fading, geometric attenuation and log-normal shadow fading and
can be expressed as \cite{marzetta10}
\begin{equation}\label{channel matrix coefficient}
{g_{minl}} = {h_{minl}}\sqrt {{\beta _{inl}}},
\end{equation}
where $h_{minl}$ is the fast fading element from the $n$th user in the $l$th cell to the $m$th antenna of the $i$th BS, which has independent real and imaginary parts with zero mean and variance $1/2$. The large scale fading $\beta_{inl}$ from the $n$th user in the $l$th cell to the $i$th BS models both the geometric attenuation and shadow fading. It is reasonable to assume the large scale fading coefficient to be constant across the antenna array, since the distance between users and the BS is much larger than the distance between antennas, and the value of $\beta_{inl}$ changes very slowly with time. Then,
\begin{equation}\label{channel matrix}
{\mathbf{G}_{0l}} = {\mathbf{H}_{0l}}{{\mathbf{D}_{0l}}^{1/2}},
\end{equation}
where $\mathbf{H}_{0l}$ denotes the $M \times N$ fast fading matrix between the users in the $l$th cell and the central cell's BS, i.e., ${[\mathbf{H}_{0l}]_{mn}} = {h_{m0nl}}$ and $\mathbf{D}_{0l}$ is a $N \times N$ diagonal matrix with ${[\mathbf{D}_{0l}]_{nn}} = {\beta_{0nl}}$.
%\subsection{Achievable Uplink Rate with ZF Receivers}
We assume all BSs have perfect channel state information (CSI) of channels in their own cell, which is a reasonable approximation in environments with low or moderate mobility. Linear processing is assumed at all receivers.
Let $\mathbf{A}_{00}$ be the $M \times N$ linear receiver matrix of the central cell's BS, which depends on the channel matrix $\bG_{00}$. The received signal vector at the BS is processed
%by multiplying it with the conjugate-transpose of the linear receiver
as
\begin{equation}\label{using receiver}
{\mathbf{r}_{0}} = \mathbf{A}_{00}^H\mathbf{y}_0^{\tt ul}.
\end{equation}
%For maximum-ratio combining (MRC) receiver, the linear detection matrix is $\mathbf{A}_{00}=\bG_{00}$. Thus, the processed signal can be written as
%\begin{equation}\label{using receiver}
%{\mathbf{r}_{0}} = \bG_{00}^H\mathbf{y}_0.
%\end{equation}
For ZF receivers,
%the linear receiver matrix is given by
$\bA_{00}= \bG_{00}\left(\bG_{00}^H \bG_{00}\right)^{-1}$. Substituting this along with \eqref{system model} into \eqref{using receiver} gives
\begin{equation}
{\mathbf{r}_0} = \bOm_0^{1/2} \bx_0 + \sum_{l=1 }^L  \bA_{00}^H {\bG_{0l}\bOm_l^{1/2} \bx_l} +\bA_{00}^H {\mathbf{n}_0^{\tt ul}}.
\end{equation}
The $n$th element of $\mathbf{r}_0$ can be further expressed as
%\begin{equation}
%{r_{0n}} = \sqrt {{p_u}} \sum\limits_{l=1}^L{\bg^H_{0n0}}\bG_{0l}\bx_l + {\bg^H_{0n0}}\mathbf{n}_0,
%\end{equation}
%where $\bg_{0n0}$ is the $n$th column of $\bG_{00}$, i.e. $\bg_{0n0} \triangleq \left[g_{1in0},\ldots,g_{min0},\ldots,g_{Min0}\right]^T$. Then, we further get
\begin{equation}\label{perfect r nth}
{r_{n0}} = \sqrt {p_{\small{n0}}^{\tt ul}}{x_{n0}} + \sum_{l=1}^L  \sum_{c=1}^N \nsp \sqrt{p_{\small{cl}}^{\tt ul}}{\ba_{0n0}^H{\mathbf{g}_{0cl}}{x_{cl}}}  + \ba_{0n0}^H\mathbf{n}_0^{\tt ul},
\end{equation}
where $\ba_{0n0}$ is the $n$th column of $\bA_{00}$, $\bg_{0cl}$ is the $c$th column of $\bG_{0l}$,
%i.e. $\bg_{0cl} = \left[g_{1icl},\ldots,g_{micl},\ldots,g_{Micl}\right]^T$
and $x_{cl}$ denotes the $c$th element of the signal vector $\mathbf{x}_l~(c=1,\ldots,N; l=1,\ldots, L)$. The desired signal in \eqref{perfect r nth} is $\sqrt{p_{n0}^{\tt ul}}x_{n0}$, and the remaining terms are considered as interference and noise. {The interference-plus-noise term is modeled as Gaussian noise, which constitutes a ``worst case'' noise assumption \cite{hassibi03}. The
%By modeling the remaining terms as additive Gaussian noise independent of $x_{0n}$, we can get a lower bound on the achievable rate.
ergodic achievable uplink rate of the $n$th user in the central cell is then given as follows \cite{Ngo11}}
%\begin{equation}\label{perfect uplink rate}
%R_{{n0}}=\bbe \left\{\log_2 \left(1+\frac{p_{n0}\left|\ba^H_{0n0}\bg_{0n0}\right|^2}{ \sum\limits_{(l,c)\ne(i,n)} p_{cl}{\left|\ba^H_{0n0}\bg_{0cl}\right|^2}
%+ \left\|\ba_{0n0}\right\|^2}\right)\right\}.
%\end{equation}
%, so that ${\bf{a}}_{0n0}^H{\bf{g}}_{0c0} = {\delta _{nc}}$, where $\delta_{nc}$ equals $1$ when $n=c$ or $0$ otherwise. Substituting the ZF receiver into \eqref{perfect uplink rate} yields
\begin{equation}\label{zf uplink rate}
R_{{n0}}^{\tt ul}=\bbe \left\{\nsp \log_2 \nsp \left(1+\frac{p_{n0}^{\tt ul}}{ \sum\limits_{l=1 }^L\sum\limits_{c=1}^N p_{cl}^{\tt ul}{\left|\ba^H_{0n0}\bg_{0cl}\right|^2}
\nsp +\nsp \left\|\ba_{0n0}\right\|^2}\right)\nsp \right\}.
\end{equation}
\subsection{Downlink}
On the downlink, we assume that the transmit power sent by the BS for each user may be different. Then, the received signal by users in the central cell can be given as
\begin{equation}\label{downlink received signal}
\mathbf{y}_{0}^{\tt dl}= \bG^{T}_{00}\bB_{00}\bs{\Psi}_0^{1/2}\mathbf{q}_0+\nsp\sum_{l=1}^L \bG^T_{l0}\bB_{ll}\bs{\Psi}_l^{1/2}\mathbf{q}_l+\mathbf{n}_{0}^{\tt dl},
\end{equation}
where $\bB_{ll}~(l=1,\ldots,L)$ is the $M \times N$ precoding matrix applied at the $l$th BS, which is dependent on the channel matrix $\bG_{ll}$, $\bs{\Psi}_{l}$ is a $N \times N$ diagonal power matrix containing $\left[p_{1l}^{\tt dl},\ldots,p_{nl}^{\tt dl},\ldots,p_{Nl}^{\tt dl}\right]$ along its main diagonal, while $p_{nl}^{\tt dl}$ is the transmitted power of the $l$th BS intended for the $n$th user. Moreover, $\mathbf{q}_l \in \mathbb{C}^{N \times 1}$ is the vector of signals intended for the $N$ users in the $l$th cell and $\mathbf{n}_{0}^{\tt dl}$ contains white complex Gaussian noise with entries having zero mean and unit variance. To satisfy the power constraint at the BS, $\bB_{ll}$ is chosen such that \cite{yang13}
\begin{equation}\label{B constraint}
\bbe\left\{{\tt tr}\left(\bB_{ll}\bB_{ll}^H\right)\right\}=1.
\end{equation}
%where ${\tt tr}(\cdot)$ means the trace of a matrix.

%\subsection{Achievable Downlink Rate with ZF Precoders}
We assume users have perfect CSI and detect received signals with the optimal maximum likelihood receivers. The ZF linear precoder is given by
\begin{equation}\label{zf precoder}
  \bB_{ll}=\alpha_l \bG_{ll}^*\left(\bG_{ll}^T\bG_{ll}^*\right)^{-1},
\end{equation}
where the constant scalar $\alpha_l$ is chosen to conform to the constraint \eqref{B constraint}.
%as
%\begin{equation}\label{derive alpha}
%\alpha^2_l \bbe\left\{{\tt tr}\left(\bB_{ll}^H \bB_{ll}\right)\right\}=1,
%\end{equation}
%which means
%\begin{equation}\label{alpha general}
%\alpha_l=\sqrt{\frac{1}{\bbe\left\{{\tt tr}\left(\bB_{ll}^H \bB_{ll}\right)\right\}}}
%\end{equation}
\begin{comment}
We know that
\begin{align}\label{trace}
\bbe\left\{{\tt tr}\left(\bG_{ll}^H\bG_{ll}\right)^{-1}\right\}
&={\tt tr}\left(\bbe\left\{\left(\bG_{ll}^H\bG_{ll}\right)^{-1}\right\}\right)\notag\\
&=\sum_{n=1}^N \bbe\left\{\left[\left(\bG_{ll}^H\bG_{ll}\right)^{-1}\right]_{nn}\right\}.
\end{align}
From \eqref{mean 1/chi-square}, it can be got that
\begin{equation}\label{mean gll}
\bbe\left\{\left[\left(\bG_{ll}^H\bG_{ll}\right)^{-1}\right]_{nn}\right\}=\frac{1}{\beta_{lnl}(M-N)}.
\end{equation}
Applying it in \eqref{trace}, we can have
\begin{equation}\label{trace 1}
\bbe\left\{{\tt tr}\left(\bG_{ll}^H\bG_{ll}\right)^{-1}\right\}=\frac{1}{M-N}\sum_{n=1}^N\frac{1}{\beta_{lnl}}.
\end{equation}
Therefore, with
\begin{equation}\label{conjugate transformation}
\bbe\left\{{\tt tr}\left(\bG_{ll}^T\bG_{ll}^*\right)^{-1}\right\}=\bbe\left\{{\tt tr}\left(\bG_{ll}^H\bG_{ll}\right)^{-1}\right\}^*,
\end{equation}
the value of $\alpha_l$ can be derived as
\begin{equation}\label{alpha_l}
\alpha_l=\sqrt{\frac{M-N}{\sum_{n=1}^N\frac{1}{\beta_{lnl}}}}.
\end{equation}
\end{comment}
The substitution of \eqref{zf precoder} into \eqref{downlink received signal} yields
\begin{equation}\label{zf downlink received signal}
\mathbf{y}_{0}^{\tt dl}= \alpha_0\bs{\Psi}_0^{1/2}\mathbf{q}_0+\nsp\sum_{l=1}^L\nsp \bG^T_{l0}\bB_{ll}\bs{\Psi}_l^{1/2}\mathbf{q}_l+\mathbf{n}_{0}^{\tt dl},
\end{equation}
and the $n$th element of $\mathbf{y}_{0}^{\tt dl}$ can be further expressed as
\begin{equation}\label{nth element of yd}
y_{n0}^{\tt dl}=\alpha_0 \sqrt{p_{n0}^{\tt dl}} q_{0n}+\sum_{l=1}^L\sum_{c=1}^N\sqrt{p_{cl}^{\tt dl}}\bg_{ln0}^T\bb_{lcl}q_{lc}+n_{n0}^{\tt dl},
\end{equation}
where $\bg_{ln0}$ and $\bb_{lcl}$ represent the $n$th and $c$th column of matrix $\bG_{l0}$ and $\bB_{ll}$, respectively, $q_{lc}$ is the $c$th element of the signal vector $\mathbf{q}_l$ and $n_{n0}^{\tt dl}$ is the $n$th element of Gaussian noise vector $\mathbf{n}_{0}^{\tt dl}$. Similarly with the uplink, the ergodic achievable downlink rate of the $n$th user in the central cell can be expressed as
\begin{equation}\label{zf downlink rate}
{R}_{n0}^{\tt dl}=\bbe\left\{\log_2\left(1+\frac{\alpha^2_0 p_{n0}^{\tt dl}}{\sum\limits_{l=1}^L \sum\limits_{c=1}^N p_{cl}^{\tt dl}\left|\bg_{ln0}^T\bb_{lcl}\right|^2+1}\right)\right\}.
\end{equation}

{\subsection{Per-cell Optimization Approach}\label{subsec: per-cell optimization}
Based on the achievable rates obtained above, we wish to design a power allocation scheme to optimize the network performance. A global (joint) optimization over all cells would be desirable, but it is found to be both very challenging and not practical due to: 1) high complexity of associated iterative algorithms (based on, e.g., game theoretic ideas), which may have convergence issues as well, and 2) the unbounded and continuous (in space) deployment of BSs which makes it impossible to define an isolated cluster (set of BSs) without being subject to other (out-of-cluster) interference. Therefore,
%In other words, even with high-complexity game-theoretic algorithms (which may have convergence issues as well), in practice we do not have an isolated cluster to optimize, but rather a continuous deployment of BSs, and the cells at the edge of a cluster will be always subject to further (out-of-cluster) interference.
%Instead of a joint optimization,
we adopt a more practical but sub-optimal per-cell optimization approach which focuses on a single-cell performance while signals from other cells are regarded as constant interference. More precisely, the sum-rate of a (target) cell is optimized while the interference footprint (caused by surrounding cells) remains unaltered. This approach entails a certain (somewhat simple) level of coordination among cells, according to which power allocation is only performed by a cell when being granted (scheduled) to do so at a particular time slot. A network management entity may, for instance, grant different time slots to different cells according to some scheduling scheme, as exemplified next.

Consider power allocation being applied in an scheduled manner to a set of cells. Cells that do not interfere with each other can perform power allocation simultaneously. Therefore,
%according to the minimum distance that cells do not affect each other,
the entire set of cells can be divided into several groups, where each group consists of geographic distinct cells that do not interfere with each other and can, thus, complete power allocation simultaneously (i.e., similar to the adoption of frequency reuse patterns). Different groups are assigned with different time slots to allocate their transmit powers. Thus, within one scheduling round, each cell has operated with maximized sum rate for at least one time slot.
 %and in the remaining slots, since we have a sum-power constraint when optimising powers in one cell, the overall interference caused by this cell to other cells is subject to a maximum level. Moreover, the result of the power allocation (using water-filling algorithm) yields that most of the power will be often concentrated in the cell center, which reduces substantially the variations in the interference power as seen by other cells.
 In the remaining slots, there will be a certain rate loss due to the variations in the interference footprint as a result of other cells' optimization. In order to minimize this loss, we will introduce a sum-power constraint in the optimization problem (detailed in the next section), such that the overall interference caused by one cell to others is upper bounded. Moreover, as it will appear, the solution to the optimization problem takes the form of a water filling algorithm.\footnote{The water filling algorithm is a general power assignment algorithm for multichannel systems with a sum-power constraint, which will assign higher powers to channels with better conditions.} It is then reasonable to expect that most of the power will be often assigned to users in the cell center, which reduces substantially the variations in the interference power as seen by other cells.
 These observations are verified by the simulation results in Section \ref{sec: numeric results}, which shows that this scheduling mechanism can achieve nearly the same performance as the joint (optimal) power allocation after one scheduling round, but with a much lower complexity.

This whole mechanism consists of two procedures: planning the scheduling sequence, and designing an optimal power allocation scheme for each cell. The first problem can be solved by frequency reuse patterns, and the second problem is our focus in this paper.}
%Therefore, from both the view of theory and practice, our power allocation schemes are valid and useful.
%Note that the achievable rates given in \eqref{zf uplink rate} and \eqref{zf downlink rate} are both difficult to analyze. In the following sections, we present some tractable expressions for them.

%\section{Analysis of achievable uplink rate}\label{sec: achievable uplink rate}
\section{Uplink Power Allocation Schemes}\label{sec: uplink}
Based on the per-cell optimization approach illustrated above, in this section, we aim to find an optimal power allocation scheme to maximize the uplink sum rate in the central cell,
%We think this assumption is reasonable, since the optimal power in one cell affects the neighbouring cell's power optimization, and without this constraint, the power allocation analysis becomes a non-stationary problem.}
%However, , we assume the total transmit power in the target cell is limited, for example, in the central cell's cell, we have $\sum\limits_{c=1}^N p_{c0}^{\tt ul} \le P^{\tt ul}$, and transmit powers from other cells are regarded as constant.
which can be obtained by solving the following optimization problem:\footnote{As stated in Section \ref{subsec: per-cell optimization}, a sum-power constraint (rather than per-user) is imposed to control the interference level.}
\begin{equation}\label{exact optimal problem}
\smash{\stackrel{*}{p}}_{n0}^{\tt ul}=\arg ~~\mathop {\max} \limits_{\sum\limits_{c=1}^N p_{c0}^{\tt ul} \le P^{\tt ul}} ~~\sum\limits_{c=1}^N R_{c 0}^{\tt ul},
%\footnote{We prefer a sum-power constraint for the optimization rather than a per-user power constraint, since the latter will lead the result of the optimization to that each user simply transmits at the max power, and no power allocation scheme would be needed. Moreover, the sum-power constraint can make the overall interference caused by this cell to other cells is subject to a maximum level, and due to the water-filling algorithm, most of the sum power will be often concentrated in the cell center, which reduces substantially the variations in the interference power as seen by other cells. All of these can support the effectiveness of the scheduled power allocation mechanism.}
\end{equation}
where $\smash{\stackrel{*}{p}}_{n0}^{\tt ul}$ denotes the optimal $p_{n0}^{\tt ul}$, {and the power allocated to each user
is adjusted at the time scale of slow fading.}
%To solve this problem, we replace $R_{c0}$ with $\tilde R_{c0}$ given by {\it Theorem \ref{theorem 1}}.
Unfortunately, it is difficult to derive exact solutions for this optimization problem with the expectation in \eqref{zf uplink rate}. We note that a closed-form expression for the achievable uplink rate with ZF receivers has been given in \cite{ngo11uplinkPerformance}, however, this expression is rather involved and does not readily facilitate the design of an optimized power allocation policy.
Instead, we derive tractable lower and upper bounds for the
achievable uplink rate, as well as an approximation which lies between these two bounds. Based on these three expressions, we then present their corresponding power allocation strategies obtained as solutions to \eqref{exact optimal problem}, which are simple and easy to compute. First, we consider the lower bound.

\subsection{Lower Bound}
\subsubsection{Closed-form Lower Bound Expression}
\begin{theorem}\label{theorem 1}
The achievable uplink rate of the $n$th user in the central cell is lower bounded as
\begin{equation}\label{uplink lower bound}
R_{n0}^{\tt ul} \ge  R^{\tt ul,L}_{n0}=\log_2 \left(1+\frac{p_{n0}^{\tt ul}\beta_{0n0}(M-N)}{\sum\limits_{l=1 }^L \sum\limits_{c=1}^N p_{cl}^{\tt ul} \beta_{0cl}+1}\right).
\end{equation}
\end{theorem}
\proof See Appendix \ref{sec: proof of theorem 1}.
\endproof

This lower bound is much simpler than the original uplink rate and facilitates the design of the following power allocation strategy.
%The upper bound for the achievable uplink rate can also be derived, which will be given in the following subsection.

\subsubsection{Power Allocation based on Lower Bound}
With \eqref{uplink lower bound}, the optimization problem \eqref{exact optimal problem} can be simplified by replacing the exact uplink rate $R_{c0}^{\tt ul}$ with the lower bound $R_{c0}^{\tt ul,L}$ as follows
\begin{equation}\label{lower bound optimal problem}
 p_{n0}^{\tt ul}=\arg~~ \mathop {\max} \limits_{\sum\limits_{c=1}^N p_{c0}^{\tt ul} \le P^{\tt ul}}~~ \sum\limits_{c=1}^N  R_{c0}^{\tt ul,L}.
\end{equation}
The following theorem gives the solution to this problem.
\begin{theorem}\label{theorem 2}
%In multicell systems, we shows an effective power allocation scheme to maximize the lower bound of uplink sum rate per cell as
The solution to the uplink power allocation problem \eqref{lower bound optimal problem} is\footnote{Observe the water-filling form of the solution. On the uplink, users get their transmit power information through the feedback from the BS, which acquires the channel information (large-scale fading) for all users and determines the uplink transmit powers. Here, the user
scheduling involving fairness is not considered, but will be considered for future work.}
%The power allocation scheme that maximizes the uplink per-cell sum rate lower bound \eqref{uplink lower bound} is
\begin{equation}\label{lower bound power allocation}
p_{n0}^{\tt ul}=\left(\mu_0^{\tt ul}-\frac{1}{d_{n0}}\right)^+,
\end{equation}
where $\mu_0^{\tt ul}$ is chosen to satisfy $\sum\nolimits_{n=1}^N p_{n0}^{\tt ul}=P^{\tt ul}$ and $d_{n0} \triangleq {\beta_{0n0}(M-N)}/{\left(\sum\nolimits_{l=1 }^L\sum\nolimits_{c=1}^N p_{cl}^{\tt ul}\beta_{0cl}+1\right)}$.
\end{theorem}
\proof
To maximize the sum rate in \eqref{lower bound optimal problem}, it is obvious that the total power needs to be set to the largest value $P^{\tt ul}$ and each user's power should be nonnegative. Using the Lagrange multiplier approach associated with Karush-Kuhn-Tucker (KKT) conditions,
% we define
%\begin{equation}\label{lagrange function}
%f(p_{10}^{\tt ul},p_{20}^{\tt ul},\ldots,p_{N0}^{\tt ul},\kappa)=\sum\limits_{c=1}^N R_{c0}^{\tt ul,L}+\kappa\left(\sum\limits_{c=1}^N p_{c0}^{\tt ul}- P^{\tt ul}\right),
%\end{equation}
%where $\kappa$ is the Lagrange multiplier. Then, by solving the following set of equations
%\begin{equation}\label{lagrange set equations}
%\left\{
%\begin{aligned}
%&\frac{\partial f(p_{10}^{\tt ul},p_{20}^{\tt ul},\ldots,p_{N0}^{\tt ul},\kappa)}{\partial p_{10}^{\tt ul}} \le 0\\
%&p_{10}^{\tt ul}\ge 0\\
% &p_{10}^{\tt ul}\frac{\partial f(p_{10}^{\tt ul},p_{20}^{\tt ul},\ldots,p_{N0}^{\tt ul},\kappa)}{\partial p_{10}^{\tt ul}}=0\\[4pt]
%&~~~~\LargerVdot \\[4pt]
%&\frac{\partial f(p_{10}^{\tt ul},p_{20}^{\tt ul},\ldots,p_{N0}^{\tt ul},\kappa)}{\partial p_{N0}^{\tt ul}} \le 0\\
%& p_{N0}^{\tt ul}\ge 0\\
%& p_{N0}^{\tt ul}\frac{\partial f(p_{10}^{\tt ul},p_{20}^{\tt ul},\ldots,p_{N0}^{\tt ul},\kappa)}{\partial p_{N0}^{\tt ul}}=0\\
%&\sum\limits_{c=1}^N p_{c0}^{\tt ul}= P^{\tt ul},
%\end{aligned}
%\right.
%\end{equation}
we can obtain the solution to \eqref{lower bound optimal problem}, which yields \eqref{lower bound power allocation}.
% Then, the power allocation scheme in \eqref{power allocation} can be obtained by using the Lagrange Multiplier Approach.
\endproof

Note that this power allocation is related to the number of BS antennas, the number of users and the large-scale fading coefficients. In particular, as $M \to \infty$, $1/d_{n0} \to 0$, and the differences in the allocated power among users vanish, meaning that this allocation strategy tends to an equal power assignment with asymptotically large $M$.
%This is not surprising,
A similar conclusion given in \cite{li10} shows that as the signal-to-noise ratio (SNR) grows, the power allocation that maximizes the sum rate converges to an equal power assignment. In our case, this result is obtained by keeping the total SNR fixed, but increasing the number of BS antennas. These observations indicate the equivalence between the two asymptotic regimes, large number of BS antennas and high SNR,
%a imply that the large number of BS antennas has the same effect as high SNR. Therefore, it can be used to substitute for increasing transmit powers,
%i.e., the equivalence between the large number of BS antennas and the high transmit power,
which is consistent with the power-scaling results given in \cite{Ngo11}.

%Next, we particularize our results to the case of a single cell with an $M$-antenna BS and $N$ users, where we omit the cell index subscripts.
%By setting $L=0$, the lower bound \eqref{uplink lower bound} reduces to
%\begin{equation}\label{single lower bound}
%R^{\tt ul}_{n} \ge R^{\tt ul,L}_{n}=\log_2 \left(1+{p_{n}^{\tt ul}\beta_{n}(M-N)}\right),
%\end{equation}
%where $R^{\tt ul}_{n}$ denotes the achievable uplink rate of the $n$th user. Based on this, the uplink sum rate over the cell is maximized by
%\begin{equation}\label{single power allocation}
%$p_{n}^{\tt ul}=\left(\mu^{\tt ul}-\frac{1}{d_n}\right)^+$,
%\end{equation}
%where $\mu^{\tt ul}$ is chosen to satisfy the power constraint $\sum\limits_{n=1}^N p_n^{\tt ul}=P^{\tt ul}$ and $d_n \triangleq \beta_{n}(M-N)$, while %$\beta_n$ is the large-scale fading coefficient of the $n$th user.
%\end{corollary}
%\proof
%By setting $L=1$ in \eqref{uplink lower bound} and \eqref{lower bound power allocation}, we can get the desired results.
%\endproof
%Note that this lower bound agrees with the result given in \cite{Ngo11}.
%and as the number of BS antennas increases, the power allocation in single-cell systems has the same tendency as for multicell systems.
% Next, we give the upper bound for the achievable uplink rate.

\subsection{Upper Bound}
\subsubsection{Closed-form Upper Bound Expression}
\begin{theorem}\label{theorem 3}
The achievable uplink rate of the $n$th user in the central cell is upper bounded as
\begin{small}
\begin{multline}\label{uplink upper bound}
R^{\tt ul}_{n0} \le R^{\tt ul,U}_{n0} = \log_2\mathlarger{\mathlarger{\mathlarger {\left(\right.}}}1+ p_{n0}^{\tt ul}\beta_{0n0}(M-N+1)\\
 \times \sum\limits_{h=1}^{\varrho(\mathcal{A}_0)}\nsp \sum\limits_{j=1}^{\tau_h(\mathcal{A}_0)}\left\{ \lambda_{h,j}(\mathcal{A}_0)\frac{(-1)^{j-1}\zeta^{-j}_{0 \langle h\rangle}}{(j-1)!}\right.\\
 \times\left.\left. \left[e^{\frac{1}{\zeta_{0\langle h\rangle }}}{\tt E_1}\left(\nsp\frac{1}{\zeta_{0\langle h\rangle }}\nsp \right)\nsp -\nsp \sum\limits_{m=0}^{j-2}(-1)^m\zeta_{0\langle h\rangle }^{m+1}m!\right]\right\}\right),
\end{multline}
\end{small}
\hspace{-4pt}where
\begin{equation}
{\tt E_1}\left(x\right) \triangleq \int_1^\infty \frac{e^{-xt}}{t}dt, ~~{\tt Re}(x)\ge 0,
\end{equation}
 is the exponential integral function of order $1$, $\zeta_{0k} \triangleq p_{cl}^{\tt ul}\beta_{0cl}~(l=1,\ldots,L; c=1,\ldots,N)$ with $k=N(l-1)+c$. Moreover, $\mathcal{A}_0 \triangleq \text{diag}\left(\zeta_{01},\zeta_{02},\ldots,\zeta_{0T}\right)$, while $T=NL$, $\varrho(\mathcal{A}_0)$ is the number of the distinct diagonal elements of $\mathcal{A}_0$, $\zeta_{0\langle 1\rangle }>\zeta_{0\langle 2\rangle }>\cdots>\zeta_{0\langle \varrho(\mathcal{A}_0)\rangle }$ are the distinct diagonal elements in decreasing order, $\tau_h(\mathcal{A}_0)$ is the multiplicity of $\zeta_{0\langle h \rangle}$, and $\lambda_{h,j}(\mathcal{A}_0)$ is the $(h,j)$th characteristic coefficient of $\mathcal{A}_0$ \cite{shin08}.
\end{theorem}
\proof
See Appendix \ref{sec: proof of theorem 2}.
\endproof

%Note that the upper bound is more complicated than the lower bound.
We can now use this upper bound to provide a corresponding power allocation strategy.

\subsubsection{Power Allocation based on Upper Bound}
After replacing the exact uplink rate $R_{c0}^{\tt ul}$ in \eqref{exact optimal problem} with the upper bound $R_{c0}^{\tt ul,U}$ given in \eqref{uplink upper bound}, the central cell optimization problem becomes
\begin{equation}\label{upper bound optimal problem}
 p_{n0}^{\tt ul}=\arg ~~\mathop {\max} \limits_{\sum\limits_{c=1}^N p_{c0}^{\tt ul} \le P^{\tt ul}}~~ \sum\limits_{c=1}^N  R_{c0}^{\tt ul,U},
\end{equation}
and the solution is presented in the following theorem.
\begin{theorem}\label{theorem 4}
The solution to the uplink power allocation problem \eqref{upper bound optimal problem} is
%The power allocation scheme that maximizes the uplink per-cell sum rate upper bound \eqref{uplink upper bound} is
\begin{equation}\label{upper bound power allocation}
p_{n0}^{\tt ul}=\left(\mu_0^{\tt ul}-\frac{1}{k_{n0}}\right)^+,
\end{equation}
where $\mu_0^{\tt ul}$ is chosen to satisfy $\sum\nolimits_{n=1}^N p_{n0}^{\tt ul}=P^{\tt ul}$ and
\begin{small}
\begin{multline}\label{bn0}
k_{n0} \triangleq  \beta_{0n0}(M-N+1) \nsp \sum\limits_{h=1}^{\varrho(\mathcal{A}_0)}\nsp \sum\limits_{j=1}^{\tau_h (\mathcal{A}_0)}\nsp \left\{ \lambda_{h,j}(\mathcal{A}_0)\frac{(-1)^{j-1}\zeta^{-j}_{0 \langle h\rangle}}{(j-1)!} \right.\\ \times \left. \left[e^{\frac{1}{\zeta_{0\langle h\rangle }}}{\tt E_1}\left(\nsp\frac{1}{\zeta_{0\langle h\rangle }}\nsp \right)\nsp -\nsp \sum\limits_{m=0}^{j-2}(-1)^m\zeta_{0\langle h\rangle }^{m+1}m!\right]\right\}.
\end{multline}
\end{small}
\end{theorem}
\proof Follows the same steps as the proof of {\it Theorem \ref{theorem 2}}.

%Note that the expression of the power allocation \eqref{bn0} is complicated. But we can find that for different $n$, the value of $k_{n0}$ is only affected by $\beta_{0n0}$ and the rest of factors can be regarded as constant.
Observe that, despite the involved structure of \eqref{bn0}, the coefficients $k_{n0}$ (for different users) differ only through the factor $\beta_{0n0}$ and the remaining factors can be regarded as constant.
 Moreover,
%the factor that really maters is $\beta_{0n0}(M-N+1)$ and the left double summation associated with the interference from other cells can be regarded as constant. Therefore, it is implied that
as $M \to \infty$, $1/k_{n0} \to 0$, and the differences among users in this power allocation scheme disappear. Again, a tendency towards an equal assignment policy is observed with $M$ growing without limit, which agrees with the observation in {\it Theorem \ref{theorem 2}}.

%Let $L=1$, the multicell system will reduce to single-cell scenario. Then, we can obtain the upper bound of achievable uplink rate and the corresponding power allocation for sing-cell systems as follows.
%Next, we particularize the result in {\it Theorem \ref{theorem 4}} for single-cell systems.
%\begin{corollary}\label{corollary 2}
%By setting $L=0$, \eqref{uplink upper bound} reduces to
%\begin{equation}\label{single upper bound}
%R^{\tt ul}_{n} \le R^{\tt ul,U}_{n}=\log_2 \left(1+{p_{n}^{\tt ul}\beta_{n}(M-N+1)}\right),
%\end{equation}
%where $R^{\tt ul}_{n}$ denotes the achievable uplink rate of the $n$th user,
%and based on this, the uplink sum rate over the whole cell is maximized by
%\begin{equation}\label{single power allocation}
%$p_{n}^{\tt ul}=\left(\mu^{\tt ul}-\frac{1}{k_n}\right)^+$,
%\end{equation}
%where $\mu^{\tt ul}$ is chosen to satisfy the power constraint $\sum\limits_{n=1}^N p_n^{\tt ul}=P^{\tt ul}$ and $k_n \triangleq \beta_{n}(M-N+1)$.
%\end{corollary}
%\proof
%Setting $L=1$ in \eqref{uplink upper bound} and \eqref{upper bound power allocation} can lead to the desired result.
%\endproof
%Comparing \eqref{single lower bound} and \eqref{single upper bound}, we find that when $M \gg N$, $R_n^{\tt ul,U}$ asymptotically approaches $R_n^{\tt ul,L}$, implying that both bounds are very tight when $M$ is large enough. Further, in the ``massive MIMO regime'', i.e., $M/N \to \infty$, these bounds become asymptotically exact.

Next, we give a third tractable expression for the achievable uplink rate---a new approximation, which is proven to lie between the upper and lower bounds derived above. As shown later in this paper, our simulations demonstrate that this result is particularly accurate over a wide range of operating conditions.
%This approximation lies between the lower and upper bound got from {\it Theorem \ref{theorem 1}} and {\ref{theorem 2}} and from the numerical results in Section \ref{sec: numeric results}, we find it is very tight.
\subsection{Approximation}\label{subsec: approximation}
\subsubsection{Closed-form Approximation}
To obtain the approximation, a key useful tool is given first. To the best of our knowledge, this approximation is new and provides a useful general tool for studying ergodic capacity.
\begin{lemma}\label{lemma 1}
If $X$ and $Y$ are independent positive random variables, then
\begin{small}
\begin{equation}\label{approximation lemma}
\bbe\left\{\log_2\left(1+\frac{X}{Y}\right)\right\}  \approx \log_2\left(1+\frac{\bbe\left\{X\right\}}{\bbe\left\{Y\right\}}\right),
\end{equation}
\end{small}
\hspace{-5pt}where
%$\log_2\left(1+\frac{\bbe\left\{X\right\}}{\bbe\left\{Y\right\}}\right)$ lies between the lower and upper bounds of $\bbe\left\{\log_2\left(1+\frac{X}{Y}\right)\right\}$ as follows
\begin{small}
\begin{equation}\label{jensen approximation lemma}
\log_2  \nmsp \left( \nsp 1 \nsp + \nsp \frac{1}{\bbe\left\{\frac{Y}{X}\right\}} \right) \le \log_2 \nmsp \left( \nmsp 1 \nsp + \nsp \frac{\bbe\left\{X\right\}}{\bbe\left\{Y\right\}} \right)  \le \log_2 \nmsp \left( \nmsp 1  \nsp + \nsp \bbe\left\{\frac{X}{Y}\right\}\nmsp \right).
\end{equation}
\end{small}
\end{lemma}
\proof
See Appendix \ref{sec: proof of lemma 1}.
\endproof

Note that the approximation in this lemma lies between two bounds obtained by a direct application of Jensen's inequality. Therefore, we can conclude that $\log_2\left(1+\frac{\bbe\left\{X\right\}}{\bbe\left\{Y\right\}}\right)$ is a relatively tight approximation of $\bbe\left\{\log_2\left(1+\frac{X}{Y}\right)\right\}$ (at least tighter than one of the two bounds). It is also important to note that the lower and upper bounds given respectively in {\it Theorems \ref{theorem 1}} and {\it \ref{theorem 3}} are obtained by applying the corresponding bounds in \eqref{jensen approximation lemma}.
%the original complicated expectation of a logarithm turns into a ratio of expectations of single random variables, which reduces the computation complexity a lot.
Moreover, this approximation becomes asymptotically exact as shown in the following corollary.
%\begin{corollary}\label{corollary 1}
%Let $X$ and $Y$ in {\it Lemma \ref{lemma 1}} be given as
%\begin{equation}\label{X and Y sum}
%X=\sum\limits_{0=1}^{\theta_1}x_0,~Y=\sum\limits_{j=1}^{\theta_2}y_j,
%\end{equation}
%where $x_0~(i=1,\ldots,\theta_1)$ and $y_j~(j=1,\ldots,\theta_2)$ are both sequences of positive square-integrable random variables which are not necessarily independent, and $\exists \epsilon>0$, $\bbe\left\{y_j\right\}> \epsilon$. Moreover, $\theta_2=c_1 \theta_1 +c_2$, whilst $c_1$ and $c_2$ are positive integers. Then, the approximation in \eqref{approximation lemma} becomes asymptotically exact as $\theta_1 \to \infty$.
%\end{corollary}
%\proof
%See Appendix \ref{sec: proof of corollary 1}.
%\endproof
%When {\it Lemma \ref{lemma 1}} is applied to the achievable rate analysis, $X$ denotes the desired signal, $Y$ is the interference-plus-noise and $\theta_1$ is the number of BS antennas. Therefore, according to {\it Corollary \ref{corollary 1}}, the approximation \eqref{approximation lemma} will be tighter as the number of BS antennas increases and becomes asymptotically exact as $M \to \infty$.

\begin{corollary}\label{corollary 2}
Let $X$ and $Y$ in {\it Lemma \ref{lemma 1}} be given as
\begin{equation}\label{X and Y sum 2}
X=1/\sum\nolimits_{i=1}^{\theta_1}x_i,~Y=\sum\nolimits_{j=1}^{\theta_2}y_j,
\end{equation}
where $x_i~(i=1,\ldots,\theta_1)$ and $y_j~(j=1,\ldots,\theta_2)$ are both sequences of positive square-integrable random variables which are not necessarily independent across $i$ and $j$, respectively, and as $\theta_1 \to \infty$, $\frac{\theta_2}{\theta_1} \to \omega \in (1,\infty)$. Moreover, there exist $\epsilon_1>0$ and $\epsilon_2>0$, such that
\begin{equation}\label{mean X Y greater 0}
\liminf_i \bbe\left\{x_i\right\}> \epsilon_1,~\liminf_j \bbe\left\{y_j\right\}> \epsilon_2.
\end{equation}
Then, as $\theta_1 \to \infty$,
\begin{small}
\begin{equation}\label{approximation asym exact}
\bbe\left\{\log_2\left(1+\frac{X}{Y}\right)\right\}  -\log_2\left(1+\frac{\bbe\left\{X\right\}}{\bbe\left\{Y\right\}}\right) \xrightarrow{a.s.} 0.
\end{equation}
\end{small}
\end{corollary}
\proof
See Appendix \ref{sec: proof of corollary 2}.
\endproof
When {\it Lemma \ref{lemma 1}} is applied to the achievable rate analysis, $\theta_1$ denotes the number of BS antennas. Therefore, according to {\it Corollary \ref{corollary 2}}, the approximation \eqref{approximation lemma} will be tighter as the number of BS antennas increases and becomes asymptotically exact as $M \to \infty$.
Note that a similar corollary can be easily obtained to show the asymptotic exactness of \eqref{approximation lemma} for the case $X=\sum\nolimits_{i=1}^{\theta_1}x_i$, which becomes useful when analyzing maximum ratio combining receivers.

Based on these results, the following theorem gives a tractable expression for the achievable uplink rate in \eqref{zf uplink rate}.
\begin{theorem}\label{theorem 5}
 The achievable uplink rate for the $n$th user in the central cell is approximated as
% \begin{small}
\begin{equation}\label{approximation}
R_{n0}^{\tt ul} \approx \tilde R_{n0}^{\tt ul}=\log_2 \left(1+\frac{p_{n0}^{\tt ul}\beta_{0n0}(M-N+1)}{\sum\limits_{l=1 }^L \sum\limits_{c=1}^N p_{cl}^{\tt ul} \beta_{0cl}+1}\right).
\end{equation}
%\end{small}
\end{theorem}
\proof
Divide the numerator and denominator of \eqref{zf uplink rate} by $\left\|\ba_{0n0}\right\|^2$. Then,
%\begin{small}
\begin{equation}\label{zf uplink rate fraction}
R_{{n0}}^{\tt ul}=\bbe \left\{\log_2 \left(1+\frac{p_{n0}^{\tt ul}\beta_{0n0}z_{0n0}}{ \sum\limits_{l=1 }^L\sum\limits_{c=1}^N p_{cl}^{\tt ul}\left|\xi_{0cl}\right|^2
+ 1}\right)\right\}.
\end{equation}
From the proof of {\it Theorem \ref{theorem 1}}, we know that $z_{0n0}$ and $\xi_{0cl}$ are independent. Hence, applying {\it Lemma \ref{lemma 1}} in \eqref{zf uplink rate fraction} yields
\begin{equation}\label{apply lemma 1}
R_{{n0}}^{\tt ul} \approx \tilde {R}_{n0}^{\tt ul}=\log_2 \left(1+\frac{p_{n0}^{\tt ul}\beta_{0n0}\bbe\left\{z_{0n0}\right\}}{ \sum\limits_{l=1 }^L\sum\limits_{c=1}^N p_{cl}^{\tt ul}\bbe\left\{\left|\xi_{0cl}\right|^2\right\}
+ 1}\right).
\end{equation}
Plugging \eqref{var gicl} and \eqref{mean 1/ain0} into \eqref{apply lemma 1} leads to the desired result.
\endproof

%Note that when $M \gg N$, the approximation in \eqref{approximation} is asymptotically equal to the lower bound given in \eqref{uplink lower bound}. Moreover,
Note from {\it Lemma \ref{lemma 1}} that the approximation in \eqref{approximation} lies between the lower and upper bounds in \eqref{uplink lower bound} and \eqref{uplink upper bound}, and as shown later in the numerical discussion, this approximation is more accurate than the aforementioned bounds. Moreover, according to {\it Corollary \ref{corollary 2}}, this approximation becomes progressively more accurate as the number of BS antennas increases. Next, we will use it for the power allocation analysis.
%\begin{figure}
%\centering{\includegraphics[width=60mm]{imagefile2.eps}}
%\caption{Quantum dot resistance for $T\ll T_{K}$ and $T\gg T_{K}$
%\source{For  high temperatures (dashed line) the Coulomb blockade remains}
%\source{For lower temperatures (solid line) the Coulomb blockade is overcome}}\label{kondodotresistance}
%\end{figure}

\subsubsection{Power Allocation based on Approximation}
\begin{comment}
Now, we assume the total power transmitted by users in one cell is limited, for example, in the central cell's cell, we have $\sum\limits_{c=1}^N p_{c0}^{\tt ul} \le P^{\tt ul}$. Under this constraint, we now aim to find an optimal power allocation scheme to maximize the sum rate per cell. That is, we need to solve the optimal problem as follows
\begin{equation}\label{optimal problem}
\smash{\stackrel{*}{p}}_{n0}^{\tt ul}=\arg \mathop {\max} \limits_{\sum\limits_{c=1}^N p_{c0}^{\tt ul} \le P^{\tt ul}} \sum\limits_{c=1}^N R_{c0}^{\tt ul},
\end{equation}
where $\smash{\stackrel{*}{p}}_{n0}^{\tt ul}$ denotes the optimal $p_{n0}^{\tt ul}$.
%To solve this problem, we replace $R_{c0}$ with $\tilde R_{c0}$ given by {\it Theorem \ref{theorem 1}}.
Unfortunately, it is difficult to derive exact closed-form solutions with the expectation in \eqref{zf uplink rate}. Here we present the
new simple analytical power allocation scheme, based on maximizing the relatively tight approximation derived in \eqref{approximation}, which is given as the following theorem.
\end{comment}
With the tractable approximation \eqref{approximation}, the central cell optimization problem becomes
\begin{equation}\label{approximation optimal problem}
 p_{n0}^{\tt ul}=\arg~~ \mathop {\max} \limits_{\sum\limits_{c=1}^N p_{c0}^{\tt ul} \le P^{\tt ul}} ~~\sum\limits_{c=1}^N \tilde R_{c0}^{\tt ul},
\end{equation}
and the following theorem presents the solution.
\begin{theorem}\label{theorem 6}
The solution to the uplink power allocation problem \eqref{approximation optimal problem} is
\begin{equation}\label{power allocation}
p_{n0}^{\tt ul}=\left(\mu_0^{\tt ul}-\frac{1}{t_{n0}}\right)^+,
\end{equation}
where $\mu_0^{\tt ul}$ satisfies $\sum\nolimits_{n=1}^N p_{n0}^{\tt ul}=P^{\tt ul}$ and $t_{n0} \triangleq {\beta_{0n0}(M-N+1)}/{\left(\sum\nolimits_{l=1 }^L\sum\nolimits_{c=1}^N p_{cl}^{\tt ul}\beta_{0cl}+1\right)}$.
\end{theorem}
\proof
Follows the same steps as the proof of {\it Theorem {\ref {theorem 2}}}.
\endproof

When $M \gg N$, the approximation in \eqref{approximation} approaches the lower bound in \eqref{uplink lower bound} asymptotically (as $M-N \to \infty$) . Therefore, for $M \gg N$, the power allocation scheme based on the approximation is also equivalent to that based on the lower bound. Hence, the power allocation strategy in {\it Theorem \ref{theorem 6}} also tends to an equal power assignment as $M \to \infty$.
%This is not surprising,

%Next, we particularize the result in {\it Theorem \ref{theorem 6}} for single-cell systems. By setting $L = 0$, the approximation \eqref{approximation} reduces to
%\begin{corollary}\label{corollary 3}
%For the case of single cell, the approximation \eqref{approximation} reduces to
%\begin{equation}\label{single approximation}
%R^{\tt ul}_{n} \approx \tilde R^{\tt ul}_{n}=\log_2 \left(1+{p_{n}^{\tt ul}\beta_{n}(M-N+1)}\right),
%\end{equation}
%and the uplink sum rate over the whole cell is maximized by
%\begin{equation}\label{single power allocation}
%$p_{n}^{\tt ul}=\left(\mu^{\tt ul}-\frac{1}{t_n}\right)^+$,
%\end{equation}
%where $\mu^{\tt ul}$ is chosen to satisfy the power constraint $\sum\limits_{n=1}^N p_n^{\tt ul}=P^{\tt ul}$ and $t_n \triangleq \beta_{n}(M-N+1)$.
%\end{corollary}
%It is found that this single-cell approximation coincides with the upper bound \eqref{single upper bound}.
 %and as the number of BS antennas increases, the power allocation in single-cell systems has the same tendency as in multicell systems.

%Now, we have given three power allocation schemes for the uplink transmission and we can deduce that since the approximation lies between the lower and upper bounds, the optimal allocation scheme will be obtained from one of the bounds or the approximation. The performance comparisons of these tractable expressions and power allocation polices are shown in Section \ref{sec: numeric results}.

 We have provided three different power allocation schemes based on three tractable expressions for the achievable rate: an upper bound, a lower bound, and an approximation which lies between these two bounds. In Section \ref{sec: numeric results}, we will evaluate the performance of these three schemes with some numerical comparisons under different operating conditions.

%\section{Analysis of Achievable Downlink Rate}\label{sec: achievable downlink rate}
\section{Downlink Power Allocation Scheme}\label{sec: downlink}

Similarly with the uplink case, we assume the total power transmitted by one BS is constrained as $\sum\nolimits_{c=1}^N p_{c0}^{\tt dl} \le P^{\tt dl}$.
Then, we aim to find an optimal power assignment
scheme to maximize the cell sum rate. That is, we
need to solve the optimization problem:
\begin{equation}\label{downlink power optimal}
\smash{\stackrel{*}{p}}_{n0}^{\tt dl}=\arg~~ \mathop {\max} \limits_{\sum\limits_{c=1}^N p_{c0}^{\tt dl} \le P^{\tt dl}} ~~\sum\limits_{c=1}^N {R}_{c0}^{\tt dl},
\end{equation}
where $\smash{\stackrel{*}{p}}_{n0}^{\tt dl}$ denotes the optimal $p_{n0}^{\tt dl}$, {and the power allocated to each user
is adjusted at the time scale of slow fading.} The exact solution of this optimal problem is hard to obtain. Therefore, aiming to obtain a simple power allocation scheme, we first present a new lower bound for the achievable downlink rate.
%In this section, we analyze the achievable downlink rate and derive a tractable lower bound, which is shown to be very tight later in numerical results. Based on it, the corresponding effective power allocation is also presented. We find that this power allocation can gain an improvement on the downlink rate performance, especially with low or moderate number of antennass and for edge users. Moreover, we also find that for downlink transmission, equal power allocation also becomes asymptotically optimal as $M \to \infty$.

\subsection{Lower Bound}
We first require the following preliminary result.
\begin{lemma}\label{lemma 2}
To satisfy the constraint \eqref{B constraint}, the constant scalar $\alpha_l$ for the ZF precoding matrix $\bB_{ll}$ is given by
%\begin{small}
\begin{equation}\label{alpha_l}
\alpha_l=\sqrt{\frac{M-N}{\sum\nolimits_{n=1}^N{1}/{\beta_{lnl}}}}.
\end{equation}
%\end{small}
\end{lemma}
\proof
See Appendix \ref{sec: proof of lemma 2}.
\endproof

We now give a tractable lower bound for the achievable downlink rate.

\begin{theorem}\label{theorem 7}
The achievable downlink rate of the $n$th user in the central cell is lower bounded by
%\begin{small}
\begin{equation}\label{zf downlink lower bound}
{R}_{n0}^{\tt dl}\ge {R}_{n0}^{\tt dl,L} = \log_2\left(1+\frac{{p_{n0}^{\tt dl}(M-N)}/{\Lambda_{00}}}{\sum\limits_{l=1 }^L\sum\limits_{c=1}^N \dfrac{p_{cl}^{\tt dl}\beta_{ln0}}{\beta_{lcl}\Lambda_{ll}}+1}\right),
\end{equation}
%\end{small}
where $\Lambda_{ll} \triangleq \sum\nolimits_{k=1}^N {1}/{\beta_{lkl}}$.
\end{theorem}
\proof
See Appendix \ref{sec: proof of theorem 7}.
\endproof

Note that this lower bound is more tractable than the exact downlink rate \eqref{zf downlink rate}. Since the ZF precoding vector $\bb_{lcl}$ in \eqref{zf downlink rate} is different across cells and users, the approach to obtain the upper bound and approximation in the uplink case does not apply here. Therefore, in the following subsection, we use this lower bound to conduct the power allocation analysis.

\subsection{Power Allocation}
With \eqref{zf downlink lower bound}, the optimization problem \eqref{downlink power optimal} can be simplified by replacing the exact uplink rate $R_{c0}^{\tt dl}$ with the lower bound $R_{c0}^{\tt dl,L}$ as
 \begin{equation}\label{downlink lower bound optimal problem}
 p_{n0}^{\tt dl}=\arg~~ \mathop {\max} \limits_{\sum\limits_{c=1}^N p_{c0}^{\tt dl} \le P^{\tt dl}}~~ \sum\limits_{c=1}^N {R}^{\tt dl,L}_{c0},
 \end{equation}
and the solution is given by the following theorem.
\begin{theorem}\label{theorem 8}
The solution to the downlink power allocation problem \eqref{downlink lower bound optimal problem} is\footnote{The channel knowledge needed in the downlink power allocation (large-scale fading coefficient) is acquired from the uplink due to channel reciprocity.}
\begin{equation}\label{downlink power allocation}
p_{n0}^{\tt dl}=\left({\mu}_0^{\tt dl}-\frac{1}{s_{n0}}\right)^+,
\end{equation}
where ${\mu}_0^{\tt dl}$ satisfies $\sum\nolimits_{n=1}^N p_{n0}^{\tt dl}=P^{\tt dl}$ and $s_{n0} \triangleq \left((M-N)/\Lambda_{00}\right)/{\left(\sum\nolimits_{l=1 }^L\sum\nolimits_{c=1}^N \frac{p_{cl}^{\tt dl}\beta_{ln0}}{\beta_{lcl}\Lambda_{ll}}+1\right)}$.
\end{theorem}
\proof
 The derivation follows the same procedure as that shown in {\it Theorem \ref{theorem 2}}.
 \endproof

 Note that the effective power allocation for the downlink is related to the number of BS antennas, the number of users and the large-scale fading coefficient. Similarly with the uplink, as $M \to \infty$, $1/s_{n0} \to 0$. Hence, the allocated power $p_{n0}^{\tt dl}$ will approach $P^{\tt dl}/N$, that is, the equal power assignment. Therefore, an important insight we can draw from comparing {\it Theorem \ref{theorem 4}} and {\it \ref{theorem 6}} is that regardless of uplink or downlink, the equal power allocation policy tends to be optimal as the number of BS antennas grows without limit. That means that large antenna arrays produce the same effect as increasing the transmit power and thus they can be used to cut down the transmit power for both uplink and downlink.

 %We also give the results for the special case of single-cell systems. By setting $L = 0$ in \eqref{zf downlink rate}, the achievable downlink rate for single-cell systems can be simplified as
  %The achievable downlink rate in the single-cell system is
 %\begin{equation}\label{downlink single lower bound}
 %{R}^{\tt dl}_n =\log_2\left(1+p_{n}^{\tt dl}(M-N)/\sum\limits_{k=1}^N\frac{1}{\beta_{k}}\right),
 %\end{equation}
 %and the downlink sum rate is maximized by
% \begin{equation}\label{downlink single power allocation}
%$p_{n}^{\tt dl}=P^{\tt dl}/N$.
%\end{equation}
% \end{corollary}

%In this case, note from \eqref{downlink single lower bound} that in downlink single-cell systems, due to the effect of the ZF precoder, there is no interference. The SNR for each user is identical and it becomes deterministic. Therefore, an equal power assignment maximizes the downlink sum rate regardless of the number of BS antennas.

\section{Numerical Results}\label{sec: numeric results}
\begin{figure}[ht]
\centering{\includegraphics[scale=1.3]{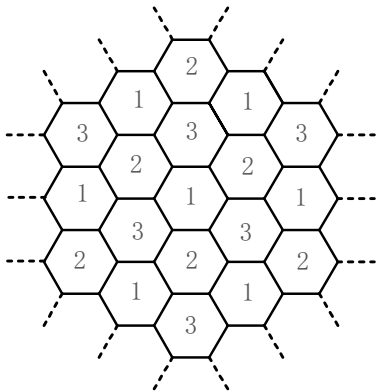}}
%[width=90mm,height=72mm]}
\caption{Simulation model.}\label{fig 1}
%\source{For  high temperatures (dashed line) the Coulomb blockade remains}
%\source{For lower temperatures (solid line) the Coulomb blockade is overcome}}\label{kondodotresistance}
\setlength{\abovecaptionskip}{0pt}
\setlength{\belowcaptionskip}{0pt}
\end{figure}

{In this section, we provide numerical results for a set of $19$ cells with radius $r_c=1000$ meters in Fig. \ref{fig 1}. $N$ users are distributed randomly and uniformly in each cell, with the exclusion of a central disk of radius $r_h=100$ meters around the BSs. The large scale fading is modeled using $\beta_{icl}={z_{icl}}/{\left({r_{icl}}/{r_h}\right)^v}$, where $z_{icl}$ is a log-normal random variable with standard deviation $\sigma$, $v$ is the path loss exponent, and $r_{icl}$ is the distance between the $c$th user in the $l$th cell and the $i$th BS. In our simulations, we choose $\sigma=8$dB and $v=3.8$. The small-scale fading is assumed to be Rayleigh distributed.} {Here, we consider a simple case where only adjacent cells produce interference (while further cells are neglected). Then, according to the scheduling approach described in Section \ref{subsec: per-cell optimization}, the whole set of cells can be divided into three groups, marked as $1$, $2$ and $3$ in Fig. \ref{fig 1}, and each group conducts power allocation in a single time slot. Since this scheduling approach is based on the per-cell optimized powers, we will first focus on the performance of a single target cell (central cell in Fig. 1), during the first time slot in which interference conditions (i.e., the powers in the other interfering cells labeled $2$ and $3$) are fixed. The network performance, taking into account dynamical effects of the interference footprint, will be considered subsequently, where we evaluate the sum rate of the entire $19$ cells
and compare our scheduled power allocation with the joint (optimal) power allocation.}

\vspace{-4ex}
\subsection{One-cell Performance}
\subsubsection{Uplink}\label{subsubsec: uplink}
For the uplink, we first evaluate the accuracy of the three expressions for the achievable rate and the performance of the power allocation schemes based on them. Then, the best power allocation strategy is selected to quantify the benefits over equal power assignment in a range of scenarios with variable $M/N$ ratios. As we will see, the approximation is the most accurate expression and the corresponding power allocation scheme can gain noticeable improvements over the equal power assignment for practical $M/N$ ratios. The transmit powers of users in surrounding $6$ cells are set to $10$dB for the uplink analysis.

%\begin{table}[ht]
%\caption{ Simulation Parameters}\label{table 1}
%%  \begin{tabularx}{80mm}{|c|>{\centering\arraybackslash}X|>{\centering\arraybackslash}X|>{\centering\arraybackslash}X|>{\centering\arraybackslash}X|>{\centering\arraybackslash}X|}
% \centering
%    \begin{tabular}{|c|>{\centering\arraybackslash}m{9mm}|>{\centering\arraybackslash}m{9mm}|>{\centering\arraybackslash}m{9mm}|>{\centering\arraybackslash}m{9mm}|>{\centering\arraybackslash}m{9mm}|}
%    \hline
%    % after \\: \hline or \cline{col1-col2} \cline{col3-col4} ...
%    Parameters & $r_c$ & $r_h$    & $\sigma$  &$v$   & $L$ \\ \hline
%    Values     & $1000$m & $100$m & $8$dB     & $3.8$ & $7$\\
%    \hline
% % \end{tabularx}
%    \end{tabular}
%\end{table}

\begin{figure}[ht]
\centering{\includegraphics[width=70mm,height=61mm]{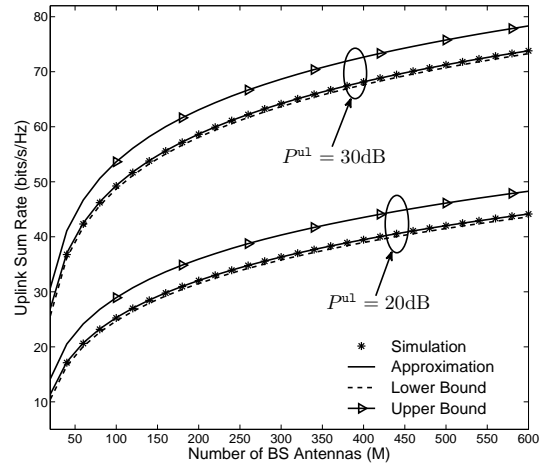}}
%[width=90mm,height=72mm]
%[width=77mm,height=61.6mm]
\caption{Uplink sum rate per cell vs. the number of BS antennas with equal power allocation, where $N=10$ users.}\label{fig 2}
%\source{For  high temperatures (dashed line) the Coulomb blockade remains}
%\source{For lower temperatures (solid line) the Coulomb blockade is overcome}}\label{kondodotresistance}
\setlength{\abovecaptionskip}{0pt}
\setlength{\belowcaptionskip}{0pt}
\end{figure}

In Fig. \ref{fig 2}, the simulated uplink sum rate in \eqref{zf uplink rate} is compared with its lower bound in \eqref{uplink lower bound}, upper bound in \eqref{uplink upper bound} and approximation in \eqref{approximation}. Results are given under two different total transmit powers---$20$ and $30$dB, and the transmit power of each user is assumed to be the same. Clearly, in all cases, the uplink rate grows with the increasing number of BS antennas. Moreover, we see a close agreement between the simulation results and our analytical lower bound and approximation. In particular, the approximation is almost indistinguishable with the simulated values and lies between the lower and upper bound, as expected from Section \ref{subsec: approximation}. Observe that the lower bound and the approximation are very close to each other; indeed, this can be anticipated from \eqref{uplink lower bound} and \eqref{approximation} when $M \gg N$.

%\end{comment}
%\vspace{1ex}

%\begin{comment}

%\end{comment}
%\vfill\pagebreak

\begin{figure}[ht]
\centering{\includegraphics[width=70mm,height=61mm]{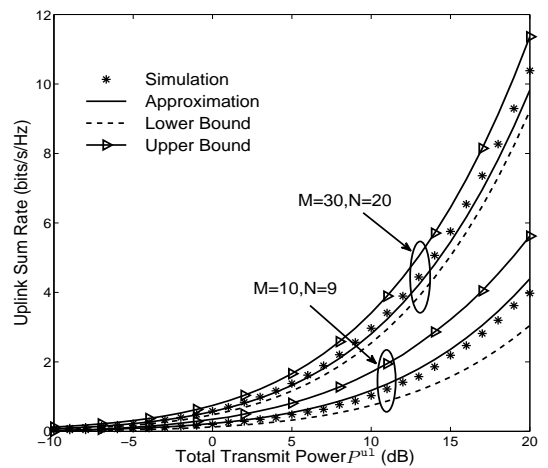}}
%[width=90mm,height=72mm]
%[width=77mm,height=61.6mm]
\caption{Uplink sum rate per cell vs. the total transmit power $P^{\tt ul}$, with equal power allocation.}\label{fig 3}
%\source{For  high temperatures (dashed line) the Coulomb blockade remains}
%\source{For lower temperatures (solid line) the Coulomb blockade is overcome}}\label{kondodotresistance}
\setlength{\abovecaptionskip}{0pt}
\setlength{\belowcaptionskip}{0pt}
\end{figure}

To show the distinction between the lower bound and approximation, we present simulation curves for $M$ closer to $N$ in Fig. \ref{fig 3}.
%In particular, comparing with the equal power assignment, the power allocation scheme given in \eqref{power allocation} leads a conspicuous improvement on the uplink performance. As expected, this improvement becomes less and less as the number of BS antennas rises, which means that for power-limited massive MIMO systems, the equal power assignment can approximately obtain the max uplink sum rate.
Here, the simulated uplink sum rate as well as the bounds and approximation are plotted against the uplink total transmit power. We can see that, with $M$ closer to $N$, the gap between the lower bound and approximation becomes more evident than that in Fig. \ref{fig 2}, and in this case, the accuracy of the upper bound is comparable to that of the lower bound. Importantly, the approximation is still the most accurate. Next, we investigate the proposed power allocation schemes.

\begin{figure}[ht]
\centering{\includegraphics[width=69mm,height=61mm]{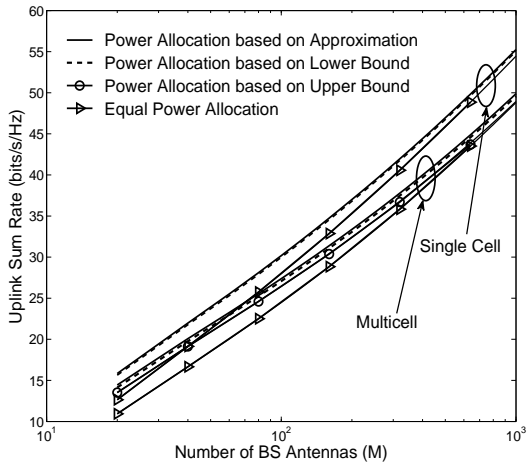}}
%[width=90mm,height=72mm]}
\caption{Uplink sum rate vs. the number of BS antennas with different power allocation schemes, where $P^{\tt ul}=20$dB and $N=10$ users.}\label{fig 4}
%\source{For  high temperatures (dashed line) the Coulomb blockade remains}
%\source{For lower temperatures (solid line) the Coulomb blockade is overcome}}\label{kondodotresistance}
\setlength{\abovecaptionskip}{0pt}
\setlength{\belowcaptionskip}{0pt}
\end{figure}

Fig. \ref{fig 4} plots the different power allocation schemes described in Section \ref{sec: uplink} under both multicell and single-cell scenarios.\footnote{The power allocation schemes in single-cell scenario are obtained by setting $L=0$. Then, the power allocation based on the approximation and that based on the upper bound have the same expression.} For multicell systems, the power allocation \eqref{lower bound power allocation} based on the lower bound, \eqref{upper bound power allocation} based on the upper bound and \eqref{power allocation} based on the approximation are compared with the equal power assignment. Clearly, all power allocation schemes achieve noticeable improvements on the uplink rate, especially when the number of BS antennas is within a few hundred. As expected, these improvements become less significant as the number of BS antennas increases,  and we observe that, when $M \to \infty$, the proposed power allocation schemes approach equal power assignments.
%i.e., the equal power allocation can approximately maximize uplink sum rate. Moreover,
It can be observed that the power allocation based on the approximation leads to the maximum enhancement, with a very slight improvement over the assignment based on the lower bound. This is a direct consequence of the increased accuracy of the approximation over the bounds.
%For single-cell systems, since the power allocation based on the approximation and that based on the upper bound have the same expression, we only use one of them to do the comparison. It is found that, compared to the equal power allocation, these power allocation schemes also have conspicuous improvements on the uplink rate, and as $M$ increases, their behaviors are in accordance with the laws in multicell systems described above.

To illustrate the effect of these power allocation schemes further, we define the {\it relative gain} of the power allocation as $\eta \triangleq \left({C_{\tt PA}-C_{\tt EQ}}\right)/{C_{\tt EQ}}$,
%\begin{equation}\label{relative gain}
%\eta \triangleq \frac{C_{\tt PA}-C_{\tt EQ}}{C_{\tt EQ}},
%\end{equation}
where $C_{\tt PA}$ denotes the sum rate with our power allocation schemes, and $C_{\tt EQ}$ denotes the sum rate under equal power allocation.

\begin{figure}[ht]
\centering{\includegraphics[width=72mm,height=61mm]{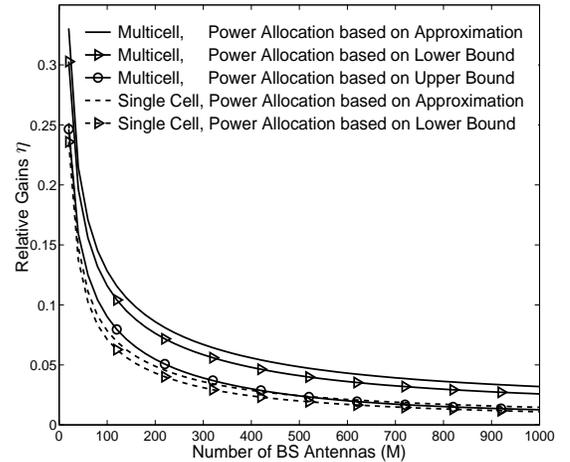}}
%[width=90mm,height=72mm]
%[width=77mm,height=61.6mm]
\caption{Relative gains of different uplink power allocation schemes, where $P^{\tt ul}=20$dB and $N=10$.}\label{fig 5}
%\source{For  high temperatures (dashed line) the Coulomb blockade remains}
%\source{For lower temperatures (solid line) the Coulomb blockade is overcome}}\label{kondodotresistance}
\setlength{\abovecaptionskip}{0pt}
\setlength{\belowcaptionskip}{0pt}
\end{figure}

Fig. \ref{fig 5} shows that, for the proposed power allocation schemes, significant gains are attained even with a few hundred antennas. For example, when $M=100$, the power allocation based on the approximation in multicell systems can obtain almost $14\%$ gain. Moreover, the gap in the multicell scenario is enhanced compared with that in the single-cell scenario. This is due to the addition of interference from other cells, which makes each user's receive SNR smaller, thus leading to a better performance of the water filling algorithm. We thus anticipate this difference to grow as the number of cells increases.

%The above analysis conclude that the gain brought by the power allocation schemes will reduce as the increase of number of BS antennas. Now, we want to figure out how much $M$ larger than $N$ can get an acceptable gain, that is, to find out when the power allocation is worth. A more general parameter, $M/N$ratio, is used to explore the effect of power allocation schemes. Due to that the power allocation based on the approximation can obtain the maximum gain, we only analyze this power allocation strategy in the following simulations.

The above results indicate that the gain brought by the power allocation schemes diminishes as the number of antennas increases while keeping the number of users fixed, i.e., as $M/N \to \infty$, which is sometimes referred to as the ``massive MIMO regime''. However, this might not be representative of practical scenarios where (i) the number of antennas is limited or (ii) the number of users is also large and is comparable with the number of antennas. In order to explore the effects of the proposed schemes in such scenarios, we next focus on the regime where $M$ and $N$ scale together with a fixed $M/N$ ratio. Since the power allocation based on the analytical approximation \eqref{power allocation} has been shown to attain the maximum gain, we only consider this strategy in the subsequent simulations.

\begin{figure}[ht]
\centering{\includegraphics[width=70mm,height=61mm]{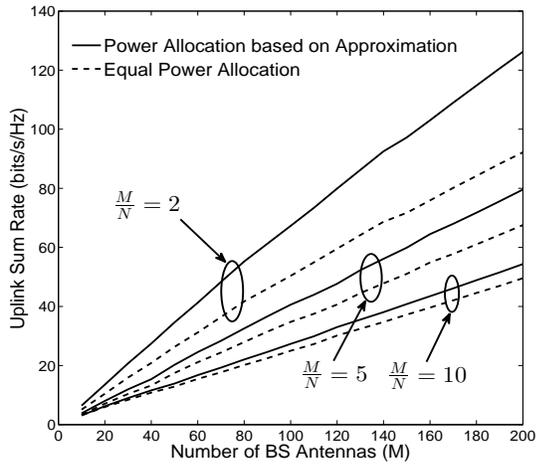}}
%[width=90mm,height=72mm]
%[width=77mm,height=61.6mm]
\caption{Uplink sum rate vs. the number of BS antennas with constant $M/N$.}\label{fig 6}
%\source{For  high temperatures (dashed line) the Coulomb blockade remains}
%\source{For lower temperatures (solid line) the Coulomb blockade is overcome}}\label{kondodotresistance}
\setlength{\abovecaptionskip}{0pt}
\setlength{\belowcaptionskip}{0pt}
\end{figure}

\begin{figure}[ht]
\centering{\includegraphics[width=70mm,height=61mm]{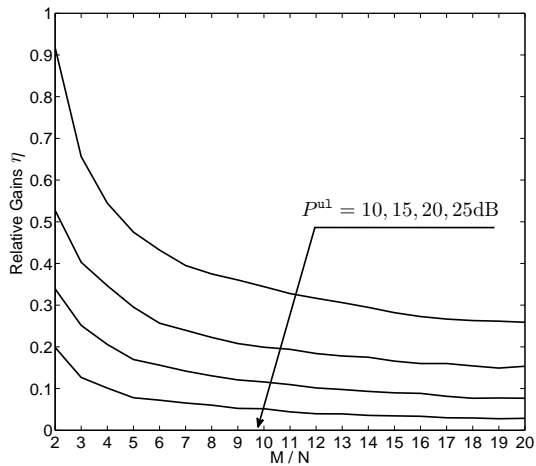}}
%[width=90mm,height=72mm]
%[width=77mm,height=61.6mm]
\caption{Relative gains of the power allocation based on approximation.}\label{fig 7}
%\source{For  high temperatures (dashed line) the Coulomb blockade remains}
%\source{For lower temperatures (solid line) the Coulomb blockade is overcome}}\label{kondodotresistance}
\setlength{\abovecaptionskip}{0pt}
\setlength{\belowcaptionskip}{0pt}
\end{figure}

Fig. \ref{fig 6} shows the effect of optimized power allocation with different $M/N$ ratios. We find that under a fixed $M/N$, the gain brought by the power allocation grows as the number of BS antennas increases, which differs from the conclusion under a fixed $N$. Since $M$ and $N$ scale together in this case, a larger number of antennas implies a larger number of users and, consequently, an increased interference which makes the receive SNR smaller. Notably, it is precisely under low SNR conditions where the power allocation optimization brings most benefit. The same arguments explain the fact that larger gains are attained at smaller $M/N$. It is also interesting to note that the curves in Fig. \ref{fig 6} are all approximately linear with no offset, which suggests that the relative rate gap between schemes with and without power allocation remains constant no matter the value of $M$. Therefore, the relative gain for each $M/N$ ratio can be obtained from the slopes of these curves via linear fitting. These results are shown in Fig. \ref{fig 7}.

Fig. \ref{fig 7} presents how the relative gains of the power allocation change with the $M/N$ ratio. We give results for four different total transmit powers---$10$, $15$, $20$ and $25$dB. As expected, the relative gains reduce with increasing $M/N$ or $P^{\tt ul}$, which is consistent with the property of the water filling algorithm. Importantly, Fig. \ref{fig 7} allows determining the $M/N$ ratios which bring gains over a particular (required) threshold, and these results can be used to decide whether the power allocation is justified. For example, if we set the tolerable minimum relative gain as $10\%$, the $M/N$ ratio with transmit power $10$ and $15$dB can be high; for example, more than $20$. This implies that even if there are more than $20$ BS antennas per user, one still obtains more than $10\%$ relative gain by using optimized power allocation. However, as the transmit power is boosted, the relative gains of power allocation will diminish. For example, for a transmit power of $20$dB, relative gains exceeding $10\%$ are only observed for $12$ BS antennas or less per user (i.e., $M/N \le 12$); whilst for a transmit power $25$dB, these gains are achieved for only $4$ BS antennas or less per user.
For system configurations in which the relative gain does not exceed a prescribed threshold (e.g., $10\%$), one may favor equal power allocation, as a consequence of its minimal complexity. This is further explored in Table \ref{table 2}, which indicates the maximum number of BS antennas per user (i.e., max value of $M/N$) required for two prescribed relative gain thresholds ($10\%$ and $20\%$), for different transmit power constraints. These results demonstrate that, even when there are $10$ times more BS antennas than users, and even for moderate transmit powers (e.g., $15$dB), one can still obtain quite substantial gains (e.g., more than $20\%$) by performing power allocation optimization.

%\vspace{-1ex}
\begin{table}[ht]
\centering
\caption{{\footnotesize Maximum $M/N$ ratios with different uplink relative gain thresholds and total transmit powers}}\label{table 2}
\vspace{1ex}
%\caption{\small Maximum $M/N$ ratios with different uplink relative gain thresholds and total transmit powers}\label{table 2}
%\normalsize
%\scriptsize
\begin{tabular}{|c|c|c|c|c|}
\hline
\multirow{2}{*}{Minimum Relative Gain} & \multicolumn{4}{c|}{$P^{\tt ul}$} \\ \cline{2-5}
                       & $10$dB    &  $15$dB   &  $20$dB   &  $25$dB   \\ \hline
               $10\%$  &   $93$  &  $30$   & $12$    &  $4$   \\ \hline
               $20\%$  &   $29$  &  $10$   &  $4$   &   $2$  \\ \hline
\end{tabular}
\end{table}

%\vspace{-2ex}

\subsubsection{Downlink}
For the downlink, we first show the performance of the lower bound for the achievable rate. Then, the corresponding power allocation scheme is evaluated for two different kinds of users---central users and edge users. We find that although the power allocation strategy still tends to an equal power assignment when $M/N \to \infty$, it obtains considerable gains with practical $M/N$ values. The transmit powers in surrounding $6$ cells are set to $30$dB for the downlink analysis.

%\vspace{2ex}

\begin{figure}[ht]
\centering{\includegraphics[width=70mm,height=61mm]{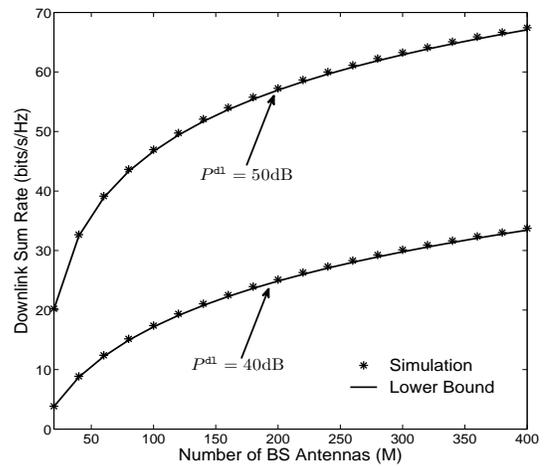}}
%[width=90mm,height=72mm]
%[width=77mm,height=61.6mm]
\figcaption{Downlink sum rate per cell vs. the number of BS antennas with equal power allocation, where $N=10$ users.}\label{fig 8}
%\source{For  high temperatures (dashed line) the Coulomb blockade remains}
%\source{For lower temperatures (solid line) the Coulomb blockade is overcome}}\label{kondodotresistance}
\setlength{\abovecaptionskip}{0pt}
\setlength{\belowcaptionskip}{0pt}
\end{figure}

In Fig. \ref{fig 8}, the simulated downlink sum rate in \eqref{zf downlink rate} is compared with the analytical lower bound in \eqref{zf downlink lower bound} under two different total transmit powers---$40$ and $50$dB. The transmit power intended for each user is assumed to be the same. Clearly, in all cases, the downlink sum rates grow as the number of BS antennas increases and we can see a close agreement between the simulation results and our analytical lower bound.

\begin{figure}
\centering{\includegraphics[width=65mm,height=61mm]{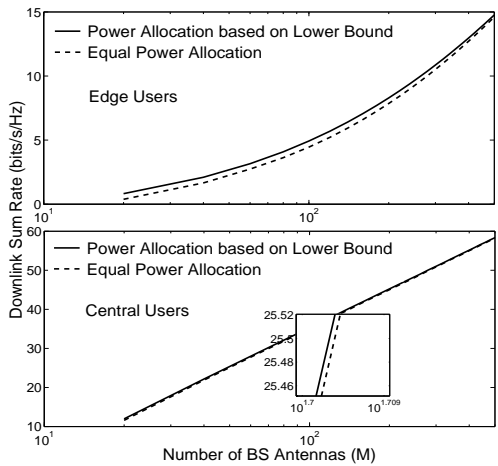}}
%[width=90mm,height=72mm]
%[width=77mm,height=61.6mm]
\caption{Downlink sum rate of central and edge users vs. the number of BS antennas with the proposed power allocation scheme, where $P^{\tt dl}=40$dB and $N=10$ users.}\label{fig 10}
%\source{For  high temperatures (dashed line) the Coulomb blockade remains}
%\source{For lower temperatures (solid line) the Coulomb blockade is overcome}}\label{kondodotresistance}
\setlength{\abovecaptionskip}{0pt}
\setlength{\belowcaptionskip}{0pt}
\end{figure}

Fig. \ref{fig 10} shows the effect of downlink power allocation based on the lower bound \eqref{downlink power allocation}. Results are given for central and edge users, which are separated by $0.8 r_c$. The power allocation scheme has a noticeable rate improvement for edge users, but for central users the improvement is almost negligible. This is because the SNR of edge users is much smaller than that of central users and, as discussed above, the gains due to power allocation are most prominent at low SNRs. In addition, when the number of BS antennas increases, the improvement brought by power allocation also reduces, and as $M \to \infty$, this power allocation scheme tends to an equal power allocation. These observations are all in accordance with our analytical results.

\begin{figure}
\centering{\includegraphics[width=68mm,height=58mm]{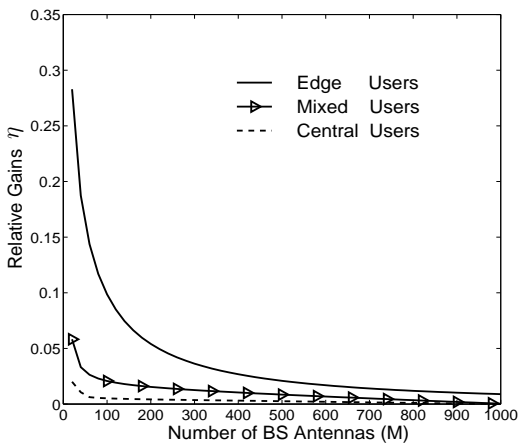}}
%[width=90mm,height=72mm]
%[width=77mm,height=61.6mm]
\caption{Relative gains of the downlink power allocation scheme, where $P^{\tt dl}=40$dB and $N=10$ users.}\label{fig 11}
%\source{For  high temperatures (dashed line) the Coulomb blockade remains}
%\source{For lower temperatures (solid line) the Coulomb blockade is overcome}}\label{kondodotresistance}
\setlength{\abovecaptionskip}{0pt}
\setlength{\belowcaptionskip}{0pt}
\end{figure}

To further illustrate the improvement due to power allocation for the downlink, Fig. \ref{fig 11} shows the relative gains of the proposed power allocation scheme over the equal power assignment. This power allocation policy provides the maximum enhancement for edge users and, as anticipated, these gains are almost negligible for central users. As expected, the gains diminish with $M$, vanishing asymptotically as $M \to \infty$, which means that the equal power allocation tends to be optimal. Note however that the gains for edge users are still noticeable for antenna numbers up to a few hundred. For example, taking $M=100$, power allocation yields a $10\%$ gain on the downlink rate for edge users, in contrast to a $2\%$ gain for central users.

%The above downlink analysis considered a fixed number of users while $M$ was allowed to grow.
Similarly to the uplink analysis, for the downlink, we are also interested in further exploring when the power allocation scheme is justified. Due to the precoding matrix constraint, the downlink sum rate does not show the linear behavior shown in Fig. \ref{fig 6} for the uplink (meaning that the sum rate depends on $M$, and not only on the ratio $M/N$). Therefore, we prescribe some specific values of $M$ and $N$ to perform this analysis, where we only consider edge users. Table \ref{table global}.a presents the maximum numbers of antennas for a minimum acceptable relative gain with different numbers of users and transmit powers. We can see that, as expected, when the total transmit power or the relative gain threshold increase, the maximum $M$ decreases. Moreover, the maximum $M$ grows as $N$ increases and this is because more users result in more interference and lower SNR, which ultimately leads to a higher relative gain from the power allocation. In Table \ref{table global}.b, we present the minimum numbers of users for an acceptable relative gain with different $M$ and transmit powers. These two tables can be used to determine whether the downlink power allocation is justified. For example, when $N=10$ with transmit power $35$dB, the power allocation guarantees a minimum relative gain of $10\%$ for any $M$ up to $288$.
But for $40$ and $45$dB, such gains are achieved for $M$ up to $99$ and $38$, respectively. Similarly, if $M$ is set to $100$, according to Table \ref{table global}.b, a system with transmit power $35$dB can obtain at least $10\%$ relative gain with more than $7$ users. But for $40$ and $45$dB, to achieve this target, there must be more than $10$ and $16$ users, respectively. As for systems that can not reach the prescribed minimum relative gain, the equal power allocation may be more suitable.

\begin{table}[ht]
\caption{ {\footnotesize Relationship between $M$ and $N$ with different downlink relative gain thresholds and total transmit powers}}\label{table global}
\vspace{1ex}
%\begin{minipage}{0.5\linewidth}
\begin{subtable}{\linewidth}
\centering

%\caption{{\scriptsize Maximum $M$ with different $N$}}\label{table 3}
%\scriptsize

\begin{tabular}{|c|c|c|c|c|}
\hline
\multirow{2}{*}{$N$}   & \multirow{2}{*}{Minimum Relative Gain}  & \multicolumn{3}{c|}{$P^{\tt dl}$} \\ \cline{3-5}
                                  &                                              & $35$dB  & $40$dB &$45$dB      \\ \hline
\multirow{2}{*}{5}    & $10\%$                                  & 48     &  18     & 9    \\ \cline{2-5}
                                & $20\%$                                  & 16      &  8     & 6     \\ \hline
\multirow{2}{*}{10}    & $10\%$                                  & 288      &  99     & 38    \\ \cline{2-5}
                                & $20\%$                                  & 89      &  36     & 18     \\ \hline
\multirow{2}{*}{20}   & $10\%$                                  & 1442      &  463    & 159     \\ \cline{2-5}
                                & $20\%$                                  & 454       &  147     & 63    \\ \hline
\end{tabular}
\vspace{2ex}
\caption{{\scriptsize Maximum $M$ with different $N$}}\label{table 3}
\vspace{1ex}
\end{subtable}
%\end{minipage}
%\end{table}
%\begin{table}[ht]
%\begin{minipage}{0.5\linewidth}
\begin{subtable}{\linewidth}
\centering
%\caption{{Minimum $N$ with different $M$}}\label{table 4}
%\scriptsize
\begin{tabular}{|c|c|c|c|c|}
\hline
\multirow{2}{*}{$M$}   & \multirow{2}{*}{Minimum Relative Gain}  & \multicolumn{3}{c|}{$P^{\tt dl}$} \\ \cline{3-5}
                       &                                         & $35$dB  & $40$dB &$45$dB      \\ \hline
\multirow{2}{*}{50}    & $10\%$                       & 5      &  7     & 12    \\ \cline{2-5}
                               & $20\%$                         & 8       &  12     & 18     \\ \hline
\multirow{2}{*}{100}   & $10\%$                      & 7      &  10    & 16     \\ \cline{2-5}
                                & $20\%$                        & 11       &  17     & 25    \\ \hline
\multirow{2}{*}{200}   & $10\%$                      & 9      &  14    & 22     \\ \cline{2-5}
                               & $20\%$                          & 14      &  25     & 36     \\ \hline
\end{tabular}
\vspace{2ex}
\caption{{\scriptsize Minimum $N$ with different $M$}}\label{table 4}
\end{subtable}
%\end{minipage}
\end{table}

{\subsection{Multicell Performance with Scheduling Mechanism}
The previous results demonstrated the performance from the perspective of a given cell, assuming that the interference power originating from the surrounding cells was fixed. This allowed us to focus on the relative pros and cons of the different power allocation strategies under ranging conditions. Now, we adopt a network perspective, and evaluate the performance of a set of cells with our scheduling mechanism for power allocation, taking the uplink as an example. We adopt the power allocation strategy based on the analytical approximation \eqref{power allocation}, as it was shown in Section \ref{subsubsec: uplink} to yield the best performance among the different schemes considered. We apply this strategy for each cell and evaluate the sum rate of the $19$ cells in Fig. \ref{fig 1}.\footnote{When evaluating the performance of these $19$ cells, some adjacent cells out of this cluster, which bring interference to cells at the edge of this cluster, should also be considered. Since the number of power allocation objectives must be finite, we assume these out-of-cluster cells have constant transmit powers.} The initial transmit power for each user is $10$dB. In time slot $i$ (for $i=1,2,3,\dots$), only cells within group $j=\bmod(i-1,3)+1$ adjust their powers. Thus, during each scheduling round (comprising $3$ time slots), each cell will operate with optimized power for one slot. In the subsequent two time slots, it will incur some performance loss due to a changing interference footprint caused by power allocation being performed in the surrounding cells. As a benchmark, we compare with the optimal performance achieved via joint power allocation across all $19$ cells,\footnote{The joint power allocation problem does not admit an analytical solution; thus, it was evaluated numerically using the function ``fmincon'' in the MATLAB optimization toolbox.} and with the simple equal power allocation strategy (i.e., no power adaptation).}

\begin{figure}
\centering
  \includegraphics[width=70mm,height=61mm]{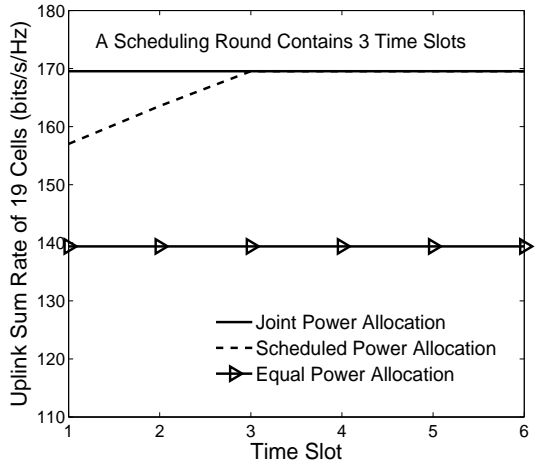}
  \caption{The uplink sum rate of $19$ cells vs. the index of time slots, where $M=20$, $N=5$ and $P^{\tt ul}=50$W.}\label{fig 12}
\end{figure}

{From Fig. \ref{fig 12}, we find that, in slots $1$ and $2$, the joint method performs better than the scheduling one. But from slot $3$, the gap between them is negligible. That is, after a single scheduling round, the scheduled power allocation achieves almost the same performance as the joint one, and this nearly optimal performance is maintained in the subsequent time slots. The joint method, highly complex and requiring high-level communication among BSs, entails a challenging implementation in practice. At a much lower complexity, the scheduled power allocation achieves nearly the same result (with differences inappreciable) after the first scheduling round which can be seen as a transient effect.  Therefore, considering both complexity and performance, the scheduling mechanism is an attractive solution.}

\section{Conclusion}\label{sec: conclusion}
We have characterized the gains in achievable rate brought by power allocation schemes in multicell MIMO systems with large antenna arrays. With ZF receivers, we have derived lower and upper bounds for the achievable uplink rate and an approximation which lies between these two bounds. With ZF precoders, we have derived a relatively simple lower bound for the achievable downlink rate. {As opposed to a joint (across all cells) optimization, we have proposed a much simpler scheduling method to plan the power allocation arrangements for different cells.} Based on this, new power allocation strategies have been presented, which can bring considerable gains over the equal power allocation assignment. These gains are substantial for moderate array sizes up to a few hundred, and
%and, under a fixed number of users, the gains diminish as $M \to \infty$ and equal power allocation then becomes optimal.
for the case where $N$ is allowed to scale with $M$, these gains increase as $M$ grows with fixed $M/N$. Based on these results, we provide applicable values of $M/N$ with a certain minimum relative gain in the achievable rates, which can be used as practical design rules to justify the use of optimized power allocation schemes over the equal assignment policy. This reveals the applicability of our allocation schemes under a wide range of scenarios with practical numbers of users and antennas. {From a network point of view, the proposed scheduling mechanism achieves almost the same performance as the joint power allocation after one scheduling round, with much reduced complexity.}

\appendices
\section{Proof of Theorem \ref{theorem 1}}\label{sec: proof of theorem 1}
Application of Jensen's inequality to \eqref{zf uplink rate} gives
\begin{small}
\begin{equation}\label{jensen lower bound}
R_{{n0}}^{\tt ul} \ge R^{\tt ul,L}_{n0} = \log_2  \left(\nsp 1 \nsp +\nsp \frac{p_{n0}^{\tt ul}}{ \sum\limits_{l=1 }^L\sum\limits_{c=1}^N p_{cl}^{\tt ul}{\bbe\left\{\left|\ba^H_{0n0}\bg_{0cl}\right|^2\right\}}
+ \bbe\left\{\left\|\ba_{0n0}\right\|^2\right\}} \nsp\right),
\end{equation}
\end{small}
\hspace{-5pt}which can be further written as
\begin{small}
\begin{equation}\label{divide and multiply}
R^{\tt ul,L}_{n0} =  \log_2  \left(\nsp 1\nsp  + \nsp\frac{p_{n0}^{\tt ul}}{ \sum\limits_{l=1 }^L \sum\limits_{c=1}^N  p_{cl}^{\tt ul}{\bbe \left\{ \left|\xi_{0cl}\right|^2\nsp\left\|\ba_{0n0}\right\|^2\right\}}
 + \bbe\left\{ \left\|\ba_{0n0}\right\|^2  \right\}} \nsp \right),
\end{equation}
\end{small}
\hspace{-5pt}where
%\begin{equation}\label{definition of Yicl}
  $\xi_{0cl}\triangleq {\ba^H_{0n0}\bg_{0cl}}/{\left\|\ba_{0n0}\right\|}$.
%\end{equation}
 Since $\bg_{0cl}$ has a rotation-invariant distribution and $\frac{\ba^H_{0n0}}{\left\|\ba_{0n0}\right\|}$ can be regarded as a column of a rotation matrix, $\xi_{0cl}$ has the same distribution as the elements of $\bg_{0cl}$ and is independent of $\ba_{0n0}$. Therefore,
\begin{equation}\label{var gicl}
\bbe\left\{\left|\xi_{0cl}\right|^2\right\}=\beta_{0cl},
\end{equation}
and \eqref{divide and multiply} can be expressed as
\begin{small}
\begin{equation}\label{rotation extract}
R^{\tt ul,L}_{n0} =\log_2 \left( 1  + \frac{p_{n0}^{\tt ul}}{\left( \sum\limits_{l=1 }^L \sum\limits_{c=1}^N p_{cl}^{\tt ul}{\bbe \left\{\left|\xi_{0cl}\right|^2\right\}}+1\right)\bbe\left\{\left\|\ba_{0n0}\right\|^2 \right\}
} \right).
\end{equation}
\end{small}
Let $z_{0n0} \triangleq \begin{small}{1}/{\left[\left(\bH_{00}^H\bH_{00}\right)^{-1}\right]_{nn}}\end{small}$, which
is chi-squared distributed with probability density \cite{gore02}
\begin{small}
\begin{equation}\label{1/zini pdf}
f({z_{0n0}}) = \frac{e^{ - {z_{0n0}}}}{{\Gamma \left( {M - N + 1} \right)}}{\left( {{z_{0n0}}} \right)^{M - N}},~z_{0n0} \ge 0,
\end{equation}
\end{small}
%\hspace{-4pt}
%so, by solving the integral $\int_0^\infty  \frac{1}{z_{0n0}}f\left( {z_{0n0}} \right)dz_{0n0}$
\hspace{-5pt}and therefore,
%\begin{equation}\label{mean 1/chi-square}
$\bbe\left\{ {1}/{z_{0n0}} \right\}={1}/({M-N})$.
%\end{equation}
Then, with
%\begin{equation}
$\left\|\ba_{0n0}\right\|^2=\left[\left(\bG_{00}^H\bG_{00}\right)^{-1}\right]_{nn}={1}/{\beta_{0n0}z_{0n0}}$,
%\end{equation}
\begin{equation}\label{mean ain0}
\bbe\left\{\left\|\ba_{0n0}\right\|^2\right\}={1}/{\beta_{0n0}(M-N)}.
\end{equation}
Plugging \eqref{var gicl} and \eqref{mean ain0} into \eqref{rotation extract} yields the result.

\section{Proof of Theorem \ref{theorem 2}}\label{sec: proof of theorem 2}
Applying Jensen's inequality in \eqref{zf uplink rate} gives
\begin{small}
\begin{equation}\label{jensen upper bound}
R_{n0}^{\tt ul} \le R^{\tt ul,U}_{n0} = \log_2  \left( 1  +\bbe\left\{ \frac{p_{n0}^{\tt ul}}{ \sum\limits_{l=1 }^L\sum\limits_{c=1}^N p_{cl}^{\tt ul}{\left|\ba^H_{0n0}\bg_{0cl}\right|^2}
 + \left\|\ba_{0n0}\right\|^2} \right\}\right).
\end{equation}
\end{small}
\hspace{-5pt}Recalling the definition of $\xi_{0cl}$ and $z_{0n0}$ in the proof of {\it Theorem \ref{theorem 1}}, and the fact that $\xi_{0cl}$ is independent of $z_{0n0}$, we obtain
%\eqref{jensen upper bound} can be further written as
%\begin{equation}\label{divide ain0}
%R^{\tt ul,U}_{n0}= \log_2  \left(1 +\bbe\left\{ \frac{{p_{n0}^{\tt ul}}{\beta_{0n0}z_{0n0}}}{ \sum\limits_{l=1 }^L \sum\limits_{c=1}^N p_{cl}^{\tt ul}{\left|\xi_{0cl}\right|^2}
%+ 1 }\right\}\right).
%\end{equation}
%Exploiting the fact that $\xi_{0cl}$ is independent of $z_{0n0}$, we obtain
\begin{small}
\begin{equation}\label{independent extract}
R^{\tt ul,U}_{n0}= \log_2  \left( \nsp 1\nsp  +\nsp  \bbe\left\{ {p_{n0}^{\tt ul}}{\beta_{0n0}z_{0n0}} \right\}\bbe\left\{ \frac{1}{ \sum\limits_{l=1 }^L \sum\limits_{c=1}^N p_{cl}^{\tt ul}{\left|\xi_{0cl}\right|^2}+1}\right\}\nsp \right).
\end{equation}
\end{small}
\hspace{-5pt}Recalling \eqref{1/zini pdf}, we can get the following expectation
%through the integral $\int_0^\infty z_{0n0}f(z_{0n0})dz_{0n0}$, which results in
%\begin{small}
\begin{equation}\label{mean 1/ain0}
\bbe\left\{{p_{n0}^{\tt ul}}{\beta_{0n0}z_{0n0}}\right\}=p_{n0}^{\tt ul}\beta_{0n0}(M-N+1).
\end{equation}
%\end{small}\par
\par Now, we focus on the remaining expectation in \eqref{independent extract}. We know that
%\begin{equation}
\begin{small}$p_{cl}^{\tt ul}\left|\xi_{0cl}\right|^2   \sim {\tt Exp}(1/p_{cl}^{\tt ul}\beta_{0cl})$\end{small},
%\end{equation}
%Let ${\tt Re}(\xi_{0cl})$ and ${\tt Im}(\xi_{0cl})$ denote the real and imaginary part of $\xi_{0cl}$, respectively. Then,
%\begin{equation}
%\phi_{0cl} \triangleq p_{cl}^{\tt ul}\left|\xi_{0cl}\right|^2=p_{cl}^{\tt ul}\left[{\tt Re}^2(\xi_{0cl})+{\tt Im}^2(\xi_{0cl})\right].
%\end{equation}
%Therefore,
%\begin{equation}\label{2 chi-square}
%\sigma_{0cl} \triangleq \frac{\phi_{0cl}}{\frac{1}{2}p_{cl}^{\tt ul}\beta_{0cl}}\sim \chi^2_2,
%\end{equation}
%where $\chi^2_2$ is the chi-square distribution with $2$ degrees of freedom. That is, $\sigma_{0cl} \sim {\tt Exp}(1/2)$, where ${\tt Exp}(1/2)$ is the exponential distribution with parameter $1/2$. Hence, we can obtain that
%\begin{equation}\label{exponential}
%\phi_{0cl} = \frac{p_{cl}^{\tt ul}\beta_{0cl}}{2}\sigma_{0cl} \sim {\tt Exp}(1/p_{cl}^{\tt ul}\beta_{0cl}).
%\end{equation}
%\begin{equation}\label{exponential}
%\end{equation}
where ${\tt Exp}(x)$ is the exponential distribution with parameter $x$.
Let
%\begin{equation}
 \begin{small}$v_0 \triangleq \sum\nolimits_{l=1 }^L \sum\nolimits_{c=1}^N p_{cl}^{\tt ul}\left|\xi_{0cl}\right|^2$\end{small}, which can be written as
 %\end{equation}
% For ease of expression, we write the double summations in $v_0$ as
a single summation, i.e.,
%\begin{equation}\label{expand v0}
$v_0=\sum\nolimits_{k=1}^{NL} \psi_{0k}$,
%\end{equation}
where $\psi_{0k} \triangleq p_{cl}^{\tt ul}\left|\xi_{0cl}\right|^2 $ with $k=N(l-1)+c$. Hence, $\psi_{0k} \sim {\tt Exp}(1/\zeta_{0k})$, where $\zeta_{0k} \triangleq p_{cl}^{\tt ul}\beta_{0cl}$ with $k=N(l-1)+c$.
%Then, we know that $v_0$ is the sum of statistically independent and not necessarily identically distributed (i.n.i.d.) exponential random variables and
The probability density function of $v_0$ is given by \cite[Theorem 2]{bletsas07}
\begin{small}
\begin{equation}\label{sum of exp pdf}
f(v_0)=\sum\limits_{h=1}^{\varrho(\mathcal{A}_0)}\sum\limits_{j=1}^{\tau_h(\mathcal{A}_0)}\lambda_{h,j}(\mathcal{A}_0)\frac{\zeta^{-j}_{0 \langle h\rangle }}{(j-1)!}v_0^{j-1}e^{-{v_0}/{\zeta_{0 \langle h\rangle }}},
\end{equation}
\end{small}
%where $\mathcal{A}_0 \triangleq \text{diag}\left(\mu_{01},\mu_{02},\ldots,\mu_{0T}\right)$, while $T=N(L-1)$, $\varrho(\mathcal{A}_0)$ is the distinct diagonal elements of $\mathcal{A}_0$, $\mu_{0\langle 1\rangle }>\mu_{0\langle 2\rangle }>\cdots>\mu_{0\langle \varrho(\mathcal{A}_0)\rangle }$ are the distinct diagonal elements in decreasing order, $\tau_h(\mathcal{A}_0)$ is the multiplicity of $\mu_{0h}$, and $\lambda_{h,j}(\mathcal{A}_0)$ is the $(h,j)$th characteristic coefficient of $\mathcal{A}_0$ \cite{shin08}.
%In \eqref{independent extract}, we need to compute
%the expectation we need to consider is $\bbe\left\{\frac{1}{v_0+1}\right\}$, which can be computed through
%\begin{equation}\label{expectation of 1/vi+1}
%\bbe\left\{\frac{1}{v_0+1}\right\}=\int_0^\infty \frac{1}{v_0+1}f(v_0) dv_0.
%\end{equation}
%The substitution of \eqref{sum of exp pdf} into \eqref{expectation of 1/vi+1} yields
\hspace{-4pt}which leads to
\begin{small}
\begin{equation}\label{substitute pdf into expectation}
\bbe\left\{\frac{1}{v_0+1}\right\}\nsp =\nsp\nsp \sum\limits_{h=1}^{\varrho(\mathcal{A}_0)}\nsp \sum\limits_{j=1}^{\tau_h(\mathcal{A}_0)}\nsp \nsp \lambda_{h,j}(\mathcal{A}_0)\frac{\zeta^{-j}_{0 \langle h\rangle }}{(j-1)!}\int_0^\infty \frac{v_0^{j-1}}{v_0\nmsp +\nmsp 1}e^{\frac{-v_0}{\zeta_{0 \langle h\rangle }}}dv_0,
\end{equation}
\end{small}
%and next, we focus on the integral in \eqref{substitute pdf into expectation}.
\hspace{-4pt}By replacing $v_0$ with $-s$,
\begin{small}
\begin{align}
\int_0^\infty \frac{v_0^{j-1}}{v_0\nmsp +\nmsp 1}e^{\frac{-v_0}{\zeta_{0 \langle h\rangle }}}dv_0%&=-\int_0^{-\infty}\frac{(-s)^{j-1}}{1-s}e^{\frac{s}{\zeta_{0\langle h\rangle }}}ds\notag\\
%&=(-1)^{j-1}\int_0^{-\infty}\frac{s^{j-1}}{s-1}e^{\frac{s}{\zeta_{0\langle h\rangle }}}ds\notag\\
&=(-1)^{j-1}\int_0^{-\infty}\frac{s^{j-1}-1+1}{s-1}e^{\frac{s}{\zeta_{0\langle h\rangle }}}ds.\label{change -x}
\end{align}
\end{small}
\hspace{-4pt}With
%\begin{equation}\label{polynominal expansion}
$s^{j-1}-1=(s-1)\left(s^{j-2}+s^{j-3}+\cdots+1\right)$,
%\end{equation}
\eqref{change -x} can be written as
\begin{small}
\begin{multline}\label{substitution polynominal expansion}
\int_0^\infty \nsp \frac{v_0^{j-1}}{v_0\nmsp +\nmsp 1}e^{\frac{-v_0}{\zeta_{0 \langle h\rangle }}}dv_0
 =(-1)^{j-1}\\
 \times \int_0^{-\infty}\nmsp \left(s^{j-2}+s^{j-3}+\cdots+1\right)e^{\frac{s}{\zeta_{0\langle h\rangle }}}ds\\
 +(-1)^{j-1}\int_0^{-\infty}\nmsp \frac{1}{s-1}e^{\frac{s}{\zeta_{0h}}}ds.
\end{multline}
\end{small}
\hspace{-7pt} Using integration by parts, we can get
\begin{small}
\begin{equation}\label{partial integration}
\int_0^{-\infty}s^{m}e^{{s}/{a}}ds=-(-1)^m a^{m+1} m!,
\end{equation}
\end{small}
\hspace{-4pt}and we also have
\begin{small}
\begin{equation}\label{E1}
  \int_0^{-\infty}\frac{1}{s-1}e^{s/a}ds=e^{1/a}{\tt E_1}(1/a).
\end{equation}
\end{small}
%where ${\tt E_1}(x)=\in\theta_1^\infty\frac{1}{t}e^{-xt}dt$ is the exponential integral function.
\hspace{-4pt}Substituting \eqref{partial integration} and \eqref{E1} into \eqref{substitution polynominal expansion} gives
\begin{small}
\begin{multline}\label{substitution two integrals}
\int_0^\infty \frac{v_0^{j-1}}{v_0\nmsp +\nmsp 1}e^{\frac{-v_0}{\zeta_{0 \langle h\rangle}}}dv_0=
(-1)^{j-1}\\
\times \left[-\sum\limits_{m=0}^{j-2}(-1)^m\zeta_{0\langle h\rangle }^{m+1}m!+e^{1/\zeta_{0\langle h\rangle }}{\tt E_1}(1/\zeta_{0\langle h\rangle })\right].
\end{multline}
\end{small}
\hspace{-5pt}The theorem follows by substituting \eqref{substitution two integrals}, \eqref{substitute pdf into expectation} and \eqref{mean 1/ain0} into \eqref{independent extract}.

\section{Proof of Lemma \ref{lemma 1}}\label{sec: proof of lemma 1}
From Jensen's inequality,
\begin{comment}
\begin{small}
\begin{equation}\label{jensen}
\log_2\nsp \left(\nsp 1\nsp + \nsp \frac{1}{\bbe\left\{\frac{Y}{X}\right\}}\nsp \right) \le \bbe\left\{\log_2 \nsp \left( \nsp 1 \nsp +\nsp \frac{X}{Y} \right) \nsp \right\}\\  \le \log_2 \nsp \left( \nsp 1\nsp +\nsp \bbe\left\{\frac{X}{Y}\right\}\nsp \right).
\end{equation}
\end{small}
\end{comment}
%\begin{comment}
\begin{small}
\begin{equation}\label{jensen}
\log_2 \left( 1 +  \frac{1}{\bbe\left\{\frac{Y}{X}\right\}} \right) \le \bbe\left\{\log_2  \left(  1  + \frac{X}{Y} \right)  \right\} \le \log_2  \left(  1 + \bbe\left\{\frac{X}{Y}\right\} \right).
\end{equation}
\end{small}
%\end{comment}
%\hspace{-4pt}
\hspace{-5pt}Since $X$ and $Y$ are independent,
\begin{small}
\begin{equation}\label{fraction jense}
\bbe\left\{\frac{X}{Y}\right\}=\bbe\left\{X\right\}\bbe\left\{\frac{1}{Y}\right\}\ge \frac{\bbe\left\{X\right\}}{\bbe\left\{Y\right\}}.
\end{equation}
\end{small}
\hspace{-5pt}Utilizing this on the left and right hand side of \eqref{jensen}, we can obtain \eqref{jensen approximation lemma}.
\begin{comment}
\begin{small}
\begin{equation}\label{jensen approximation}
\log_2 \nsp \left( \nsp 1 \nsp + \nsp \frac{1}{\bbe\left\{\frac{Y}{X}\right\}}\right) \le \log_2\nsp \left( \nsp 1 \nsp + \nsp \frac{\bbe\left\{X\right\}}{\bbe\left\{Y\right\}}\right)\\  \le \log_2 \nsp \left(\nsp 1 \nsp + \nsp \bbe\left\{\frac{X}{Y}\right\}\nmsp \right).
\end{equation}
\end{small}
\end{comment}
%\begin{comment}
%\begin{multline}\label{jensen approximation}
%\log_2  \left(  1  +  \frac{1}{\bbe\left\{\frac{Y}{X}\right\}}\right) \le \log_2 \left(  1  +  \frac{\bbe\left\{X\right\}}{\bbe\left\{Y\right\}}\right)\\  \le \log_2  \left( 1  +  \bbe\left\{\frac{X}{Y}\right\}\nmsp \right).
%\end{multline}
%\end{comment}

Comparing \eqref{jensen} and \eqref{jensen approximation lemma}, it is obvious that $\log_2\left(1+\frac{\bbe\left\{X\right\}}{\bbe\left\{Y\right\}}\right)$ lies between the lower and upper bound of $\bbe\left\{\log_2\left(1+\frac{X}{Y}\right)\right\}$. Therefore, we get the approximation expression in \eqref{approximation lemma}.

\section{Proof of Corollary \ref{corollary 2}}\label{sec: proof of corollary 2}
According to the law of large numbers,
\begin{small}
\begin{equation}\label{X large law 2}
\frac{1}{\theta_1}\sum\nolimits_{i=1}^{\theta_1} x_i - \frac{1}{\theta_1}\sum\nolimits_{i=1}^{\theta_1} \bbe\left\{x_i \right\} \xrightarrow{a.s.} 0 ,~\mbox{as }\theta_1 \to \infty.
\end{equation}
\end{small}
\hspace{-5pt}Then, since $\frac{1}{\theta_1}\sum\nolimits_{i=1}^{\theta_1}\bbe\left\{x_i\right\}$ is bounded away from $0$,
\begin{small}
\begin{equation}\label{1/X large law}
\frac{1}{\frac{1}{\theta_1}\sum\nolimits_{i=1}^{\theta_1} x_i } - \frac{1}{\frac{1}{\theta_1}\sum\nolimits_{i=1}^{\theta_1}\bbe\left\{x_i \right\}} \xrightarrow{a.s.} 0, ~\mbox{as }\theta_1 \to \infty.
\end{equation}
\end{small}
\hspace{-5pt}That is,
\begin{small}
\begin{equation}\label{mean 1/X large law}
\bbe\left\{\frac{1}{\frac{1}{\theta_1}\sum\nolimits_{i=1}^{\theta_1} x_i } \right\}- \frac{1}{\frac{1}{\theta_1}\sum\nolimits_{i=1}^{\theta_1}\bbe\left\{x_i \right\}} \xrightarrow{a.s.} 0, ~\mbox{as }\theta_1 \to \infty.
\end{equation}
\end{small}
\hspace{-5pt}Combining \eqref{1/X large law} and \eqref{mean 1/X large law}, we get
\begin{small}
\begin{equation}\label{1/X large law 2}
\frac{1}{\frac{1}{\theta_1}\sum\nolimits_{i=1}^{\theta_1} x_i } - \bbe\left\{\frac{1}{\frac{1}{\theta_1}\sum\nolimits_{i=1}^{\theta_1} x_i } \right\}\xrightarrow{a.s.} 0, ~\mbox{as }\theta_1 \to \infty,
\end{equation}
\end{small}
\hspace{-5pt}which means
\begin{equation}\label{X large law 3}
\theta_1 X -\theta_1 \bbe\left\{X\right\} \xrightarrow{a.s.} 0, ~\mbox{as }\theta_1 \to \infty.
\end{equation}
We also have
\begin{small}
\begin{equation}\label{Y large law 2}
\frac{1}{\theta_2}Y-\frac{1}{\theta_2}\bbe\left\{Y\right\}\xrightarrow{a.s.} 0 ,~\mbox{as }\theta_2 \to \infty.
\end{equation}
\end{small}
\hspace{-5pt}Therefore, since $\frac{1}{\theta_2}\bbe\left\{Y\right\}$ is also bounded away from $0$,
\begin{small}
\begin{equation}\label{X divide by Y 2}
\frac{\theta_1 X}{\frac{1}{\theta_2}Y}-\frac{\theta_1 \bbe\left\{X\right\}}{\frac{1}{\theta_2}\bbe\left\{Y\right\}}\xrightarrow{a.s.}0,~\mbox{as }\theta_1 \to \infty ~\mbox{with }\frac{\theta_2}{\theta_1}\to \omega,
\end{equation}
\end{small}
\hspace{-5pt}which can be written as
\begin{small}
\begin{equation}\label{X divede by Y 3}
\omega\theta_1^2\left(\frac{X}{Y}-\frac{\bbe\left\{X\right\}}{\bbe\left\{Y\right\}}\right)\xrightarrow{a.s.}0,~\mbox{as }\theta_1 \to \infty.
\end{equation}
\end{small}
\hspace{-5pt}By recalling the definition of almost sure convergence, we know that $\forall \varepsilon>0$, there exists $\theta_0$, such that for $\theta_1 > \theta_0$,
\begin{small}
\begin{equation}\label{definition of almost sure 2}
{ \tt Pr}\left(\left|\omega\theta_1^2\left(\frac{X}{Y}-\frac{\bbe\left\{X\right\}}{\bbe\left\{Y\right\}}\right)\right| < \varepsilon \right)=1,
\end{equation}
\end{small}
\hspace{-5pt}where ${\tt Pr}\left(\cdot \right)$ denotes probability. Hence,
\begin{small}
\begin{equation}\label{definition of almost sure without cof 2}
{ \tt Pr}\left(\left|\frac{X}{Y}-\frac{\bbe\left\{X\right\}}{\bbe\left\{Y\right\}}\right| < \varepsilon \right)=1,
\end{equation}
\end{small}
\hspace{-5pt}and therefore
\begin{small}
\begin{equation}\label{corollary 2 result}
\frac{X}{Y} - \frac{\bbe\left\{X\right\}}{\bbe\left\{Y\right\}} \xrightarrow{a.s.} 0 ,~\mbox{as }\theta_1 \to \infty.
\end{equation}
\end{small}
\hspace{-5pt}Then, Corollary \ref{corollary 2} follows by a direct application of \eqref{corollary 2 result}.

\section{Proof of Lemma \ref{lemma 2}}\label{sec: proof of lemma 2}
Utilizing the property
%\begin{equation}\label{trace property}
\begin{small}${\tt tr}\left(\bB_{ll}\bB_{ll}^H\right)={\tt tr}\left(\bB_{ll}^H\bB_{ll}\right)$\end{small}
%\end{equation}
on \eqref{B constraint}, we can obtain
\begin{small}
\begin{equation}\label{trace change}
\bbe\left\{{\tt tr}\left(\bB_{ll}^H\bB_{ll}\right)\right\}=1.
\end{equation}
\end{small}
\hspace{-5pt}Then, substituting \eqref{zf precoder} into \eqref{trace change} yields
%\begin{equation}\label{derive alpha}
\begin{small}$\alpha^2_l \bbe\left\{{\tt tr}\left(\bG_{ll}^T \bG_{ll}^*\right)^{-1}\right\}=1$\end{small},
%\end{equation}
which implies
\begin{small}
\begin{equation}\label{alpha general}
\alpha_l=\sqrt{{1}/{\bbe\left\{{\tt tr}\left(\bG_{ll}^T \bG_{ll}^*\right)^{-1}\right\}}}.
\end{equation}
\end{small}
\hspace{-5pt}We know that
\begin{small}
\begin{align}\label{trace}
\bbe\left\{{\tt tr}\left(\bG_{ll}^H\bG_{ll}\right)^{-1}\right\}
&={\tt tr}\left(\bbe\left\{\left(\bG_{ll}^H\bG_{ll}\right)^{-1}\right\}\right)\\
&=\sum\limits_{n=1}^N \bbe\left\{\left[\left(\bG_{ll}^H\bG_{ll}\right)^{-1}\right]_{nn}\right\}.
\end{align}
\end{small}
\hspace{-5pt}From \eqref{mean ain0}, we have
\begin{small}
\begin{equation}\label{mean gll}
\bbe\left\{\left[\left(\bG_{ll}^H\bG_{ll}\right)^{-1}\right]_{nn}\right\}=\frac{1}{\beta_{lnl}(M-N)}.
\end{equation}
\end{small}
\hspace{-5pt}Applying this in \eqref{trace} yields
\begin{small}
\begin{equation}\label{trace 1}
\bbe\left\{{\tt tr}\left(\bG_{ll}^H\bG_{ll}\right)^{-1}\right\}=\frac{1}{M-N}\sum\limits_{n=1}^N\frac{1}{\beta_{lnl}}.
\end{equation}
\end{small}
\hspace{-5pt}Therefore, since \begin{small}${\tt tr}\left(\bG_{ll}^T\bG_{ll}^*\right)^{-1}={\tt tr}\left(\bG_{ll}^H\bG_{ll}\right)^{-1}$\end{small},
$\alpha_l$ is got by applying \eqref{trace 1} in \eqref{alpha general}.

\section{Proof of Theorem \ref{theorem 7}}\label{sec: proof of theorem 7}
Applying Jensen's inequality in \eqref{zf downlink rate} yields
\begin{small}
\begin{equation}\label{jensen lower bound downlink}
{R}_{n0}^{\tt dl}\ge {R}_{n0}^{\tt dl,L}= \log_2\left( 1+\frac{\alpha^2_0 p_{n0}^{\tt dl}}{\sum\limits_{l=1 }^L \sum\limits_{c=1}^N  p_{cl}^{\tt dl}\bbe\left\{\left|\bg_{ln0}^T\bb_{lcl}\right|^2\right\}+1} \right),
\end{equation}
\end{small}
\hspace{-5pt}which can be further written as
\begin{small}
\begin{equation}\label{downlink diveide and multiply}
{R}_{n0}^{\tt dl,L}= \log_2\nsp\left(\nsp 1+\frac{\alpha^2_0 p_{n0}^{\tt dl}}{\sum\limits_{l=1 }^L \nsp\sum\limits_{c=1}^N \nsp p_{cl}^{\tt dl}\bbe\left\{\nsp\frac{\left|\bg_{ln0}^T\bb_{lcl}\right|^2}{\left\|\bb_{lcl}\right\|^2}\left\|\bb_{lcl}\right\|^2\nsp \right\}+1}\nmsp \right).
\end{equation}
\end{small}
\hspace{-5pt}From the proof of {\it Theorem \ref{theorem 1}}, we know that $\frac{\bg_{ln0}^T\bb_{lcl}}{\left\|\bb_{lcl}\right\|}$ has the same distribution as $\bg_{0n0}$ and is independent of $\bb_{lcl}$. Therefore
\begin{small}
\begin{equation}\label{downlink independent extract}
{R}_{n0}^{\tt dl,L}=  \log_2\left(1+\frac{\alpha^2_0 p_{n0}^{\tt dl}}{\sum\limits_{l=1 }^L \sum\limits_{c=1}^N  p_{c0}^{\tt dl}\bbe\left\{ \frac{\left|\bg_{ln0}^T\bb_{lcl}\right|^2}{\left\|\bb_{lcl}\right\|^2} \right\}\bbe\left\{ \left\|\bb_{lcl}\right\|^2 \right\}+1} \right),
\end{equation}
\end{small}
\hspace{-5pt}and
\begin{small}
\begin{equation}\label{mean gln0}
\bbe\left\{{\left|\bg_{ln0}^T\bb_{lcl}\right|^2}/{\left\|\bb_{lcl}\right\|^2}\right\}=\beta_{ln0}.
\end{equation}
\end{small}
\hspace{-5pt}With
%\begin{equation}\label{blcl}
$\left\|\bb_{lcl}\right\|^2=\alpha^2_l\left[\left(\bG_{ll}^T\bG_{ll}^*\right)^{-1}\right]_{cc}$
%\end{equation}
and \eqref{mean gll}, we have
\begin{small}
\begin{equation}\label{mean blcl}
\bbe\left\{\left\|\bb_{lcl}\right\|^2\right\}=\frac{\alpha^2_l}{\beta_{lcl}(M-N)}.
\end{equation}
\end{small}
\hspace{-5pt}The theorem then follows by substituting \eqref{alpha_l}, \eqref{mean gln0} and \eqref{mean blcl} into \eqref{downlink independent extract}.

\end{document}